\documentclass[aps,twocolumn,pra,superscriptaddress]{revtex4}
\usepackage[T1]{fontenc}
\usepackage{appendix}
\usepackage{todonotes}
\usepackage{amsmath,amsfonts,amssymb,color,colordvi,epsfig,graphicx}
\usepackage{mathtools}
\usepackage{hyperref}
\usepackage{physics}

\newcommand{\id}{\mathbb{I}}

\begin{document}

\title{Experimental study of quantum uncertainty from lack of information}

\author{Yuan-Yuan Zhao}
\thanks{These authors contributed equally.}
\affiliation{Key Laboratory of Quantum Information, University of Science and Technology of China, CAS, Hefei, 230026, China}
\affiliation{Center for Quantum Computing, Peng Cheng Laboratory, Shenzhen 518055, China}
\affiliation{CAS Center for Excellence in Quantum Information and Quantum Physics,
University of Science and Technology of China, Hefei, 230026, China}

\author{Filip Rozp\k{e}dek}
\thanks{These authors contributed equally.}
\affiliation{QuTech, Delft University of Technology, Lorentzweg 1, 2628 CJ Delft, The Netherlands}
\affiliation{Kavli Institute of Nanoscience, Delft University of Technology, Lorentzweg 1, 2628 CJ Delft, The Netherlands}
\affiliation{Pritzker School of Molecular Engineering, University of Chicago, Chicago, IL 60637, USA}

\author{Zhibo Hou}
\author{Kang-Da Wu}

\author{Guo-Yong Xiang}
\email{gyxiang@ustc.edu.cn}

\author{Chuan-Feng Li}

\author{Guang-Can Guo}
\affiliation{Key Laboratory of Quantum Information, University of Science and Technology of China, CAS, Hefei, 230026, China}
\affiliation{CAS Center for Excellence in Quantum Information and Quantum Physics,
University of Science and Technology of China, Hefei, 230026, China}

\date{\today}

\begin{abstract}
Quantum uncertainty is a well-known property of quantum mechanics that states the impossibility of predicting measurement outcomes of multiple incompatible observables simultaneously. In contrast, the uncertainty in the classical domain comes from the lack of information about the exact state of the system. One may naturally ask, whether the quantum uncertainty is indeed a fully intrinsic property of the quantum theory, or whether similar to the classical domain lack of knowledge about specific parts of the physical system might be the source of this uncertainty. This question has been addressed in the previous literature where the authors argue that in the entropic formulation of the uncertainty principle that can be illustrated using the so-called, guessing games, indeed such lack of information has a significant contribution to the arising quantum uncertainty. Here we investigate this issue experimentally by implementing the corresponding two-dimensional and three-dimensional guessing games. Our results confirm that within the guessing-game framework, the quantum uncertainty to a large extent relies on the fact that quantum information determining the key properties of the game is stored in the degrees of freedom that remain inaccessible to the guessing party. Moreover, we offer an experimentally compact method to construct the high-dimensional Fourier gate which is a major building block for various tasks in quantum computation, quantum communication, and quantum metrology.
\end{abstract}

\maketitle

\bigskip

\section*{Introduction}

In classical physics, one can predict the outcomes of simultaneous measurements of various observables performed on the same physical system with arbitrary precision, provided that one is in possession of measuring devices that allow for reaching sufficiently high accuracy. However, the quantum theory imposes intrinsic limitations on one's ability to make such measurement predictions for the incompatible observables. The first statement which quantified this quantum uncertainty was originally proposed by Heisenberg~\cite{1927} and then rigorously proven by Kennard~\cite{kennard1927quantenmechanik} in 1927. This statement applies to two maximally incompatible observables of position and momentum of a particle and the uncertainty is characterized in terms of the standard deviation. Their work was then generalized to any two bounded Hermitian observables by Robertson~\cite{robertson1929} as:
\begin{equation}
\Delta S\cdot\Delta T\geq\frac{1}{2}\abs{\bra{\psi}[S,T]\ket{\psi}} \, ,
\end{equation}
where $\Delta S$ ($\Delta T$) denotes the standard deviation of the distribution of outcomes when observable $S$ ($T$) is measured on quantum state $\ket{\psi}$.

Unfortunately, there are various shortcomings to Robertson's uncertainty relation (see e.g.~\cite{2017entropic}) of which the most notable one is that its right hand side depends on the input state. This results in the fact that one can find states $\ket{\psi}$ for which it is impossible to predict the measurement outcome of neither $S$ nor $T$ with certainty, yet the bound  becomes trivially zero when evaluated on $\ket{\psi}$. A natural way to overcome these limitations is to consider entropic formulations of the quantum uncertainty principle which allow for state-independent bounds and provide information-theoretic interpretations of the uncertainty~\cite{2017entropic}.

For rank-one projective measurements on the finite-dimensional Hilbert space, an example of such a formulation is the well-known entropic uncertainty relation due to Maassen and Uffink \cite{maassen1988},
 \begin{equation}
 H(S)+H(T)\geq \log_2\frac{1}{c},
 \label{eq:entropic}
 \end{equation}
where $H(S)$ is Shannon's entropy of the probability distribution of the outcomes when $S$ is measured and similarly for $T$. The term $c$ on the right hand side denotes the maximum overlap of the observables, that is $c = \max_{ij} \abs{\ip{s_i}{t_j}}^2$, where $\ket{s_i}$ $(\ket{t_j})$ denotes the eigenstate of $S$ $(T)$. From the inequality (\ref{eq:entropic}), we can see that the uncertainty always exists ($\log_2\frac{1}{c}\neq0$) as long as $S$ and $T$ do not share any common eigenvector. It is then natural to raise the question regarding the origin of this uncertainty, since we already know that it is not related to the precision of the measuring apparatus.

Here we experimentally investigate this question with regard to a so-called guessing game~\cite{berta2010} that provides an operational interpretation to the entropic formulation of the uncertainty principle. In such a guessing game one attempts to guess the outcome of a measurement on a state that one can freely prepare, where the measured observable is not predetermined, but is chosen uniformly at random from a set of two incompatible observables. Not only does the guessing game perspective provide us with useful insights into the foundational aspects of the uncertainty principle but it also makes the entropic formulation of this principle a useful tool for proving security of various quantum cryptographic protocols~\cite{2017entropic}. In~\cite{rozpkedek2017} the authors have shown that in this formulation of the uncertainty principle, not all of the quantum uncertainty, and in some cases even none, should be thought of as intrinsic to the quantum nature of this game. In fact it can be attributed to the guessing party's lack of \emph{quantum} information about the choice of the measured observable. Revealing this quantum information enables the guessing party to significantly decrease, and in some cases even completely eliminate the observed uncertainty.

Here we experimentally verify the main claims of~\cite{rozpkedek2017}. That is, by experimentally implementing the discussed guessing game in which the quantum information about the state of the measuring apparatus is revealed to the guessing party, we verify that the lack of access to this information is a key contributor to the arising uncertainty. Furthermore, we propose an innovative way to construct the high-dimensional quantum Fourier transform.

Fourier transform is one of the most important tools in quantum information processing, especially in quantum algorithms involving phase estimation, including the order-finding problem and the factoring problem~\cite{nielsen2011quantum}. A notable example is Shor's factoring algorithm which shows quantum advantages over its classical counterparts \cite{Shor1994}. With the applications in quantum state
tomography and quantum key distribution, quantum Fourier transforms are
usually used to generate the mutually unbiased bases for extracting more
information from the system \cite{wootters1989,miles2013,zeilinger2006}.

Since the quantum Fourier transform occupies such an important position in quantum information and computation, people explore many protocols to implement it in different physical systems, such as superconducting system \cite{FTsuperconducting}, trapped ions \cite{FTtrappedions}, photons \cite{FTAOM,FTtimebin}, and nuclear magnetic resonance systems \cite{FTNMR}.
In our work, the high-quality two-dimensional and three-dimensional Fourier transforms are implemented on the path degree of freedom (DoF) of a single photon. Then the controlled Fourier gates with the two-dimensional control system are also realized. In our experiment all the visibilities of the three interferometers used to construct the quantum Fourier gate for $d=2$ guessing game and six interferometers in the case of the $d=3$ guessing game, are higher than $0.98$. In comparison with other DoFs of the photon, e.g., the time-bin and the orbital angular momentum, the path DoF has its advantages and is much easier to control with common beam splitters and waveplates. Furthermore, the method we adopt to construct the Fourier gate may inspire other ways to manipulate the path-encoded qudits on the integrated quantum photonic device.

To construct the three-dimensional Fourier gate, we develop an experimentally friendly structure HBD-HWP-HBD, i.e., two horizontally placed beam displacers (HBDs) with a half-wave plate (HWP) inserted between them, to realize the principle component $R_y$, the single-qubit rotation gate around $y$-axis. This HBD-HWP-HBD structure eases the complexity of the the original scheme~\cite{clements2016optimal} and reduces the scale of the setup. To be specific, for the three-dimensional Fourier transform implemented in the
experiment, three interferometers are constructed instead of six ones with the 50:50 BSs. Meanwhile, the parallel distribution structure of the beams in our method enhances the stability of the experimental setup
and makes it more robust to the environmental noise.

The paper is structured as follows. In Result section we first introduce the framework of the guessing game and provide a high-level overview of our results. We then describe the experimental results in detail and discuss their implications for verifying the claims of~\cite{rozpkedek2017}. We conclude in Discussion section where we explain the implications of our results for quantum cryptography and discuss the possible extensions of the studied guessing game that could potentially be realized on a modified version of our experimental setup. Finally, in Methods section we describe our optical implementation of this game, as well as the settings of our experimental devices that allow us to prepare quantum states needed to verify the claims of the paper.

\section*{Results}
\noindent\textbf{Guessing game}
\label{sec:GuessingGame}

\begin{figure}
\begin{center}
    	\includegraphics[width=0.9\columnwidth]{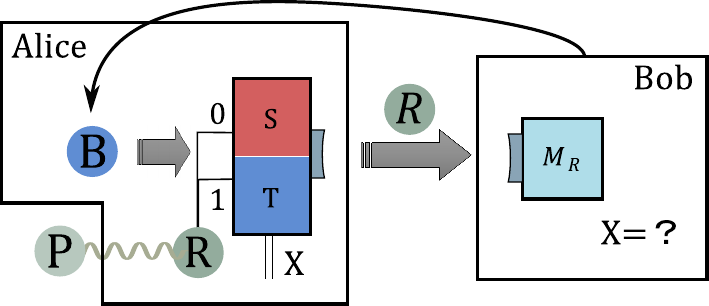}
        \caption{{\bf Guessing game.} In the $d$-dimensional guessing game, Bob prepares a quantum state $\rho_B$ of dimension $d$ and sends it to Alice. Then Alice performs the measurement $S$ or $T$ on the system $\rho_B$ according to the two-dimensional register state $\rho_R$ through a quantum control as shown in FIG.~\ref{fig:circuit}. After Alice completes the measurement, Bob tries to guess Alice's measurement outcome $X=x$ by measuring the register state $\rho_R^{x}$. In this process, $R$ can be entangled with a system $P$, which remains inaccessible to Bob. Since some information about Alice's measurement process can be contained in that register $P$, in that case Bob cannot obtain full quantum information about Alice's measurement. }\label{fig:game}
    \end{center}
\end{figure}

In this subsection we review the framework and the results of~\cite{rozpkedek2017} which form the basis for our experiment. We depict the considered guessing game (also referred to as the uncertainty game), firstly proposed by Berta \textit{et al.}~\cite{berta2010} in FIG. \ref{fig:game}. In the game, Bob prepares the system $B$ in state $\rho_B$ and sends it to Alice. Then Alice performs one of the two pre-agreed measurements $S$ and $T$ on the system according to a random coin flip contained in the two-dimensional register $R$. She announces the chosen measurement to Bob who wants to guess Alice's outcome. In particular, Bob aims to minimize his uncertainty about Alice's measurement outcome $X$ by choosing a suitable probe state $\rho_B$. The only scenario in which Bob can win the game with probability one is the game in which $S$ and $T$ share at least one common eigenvector, which corresponds to $\log_2\frac{1}{c}=0$ in the entropic uncertainty relation (\ref{eq:entropic}). In this situation, Bob prepares the probe state $\rho_B$ as the common eigenstate of $S$ and $T$, which enables him to predict the outcome of either of the measurements with certainty.

\begin{figure}
\begin{center}
\includegraphics[width=0.9\columnwidth]{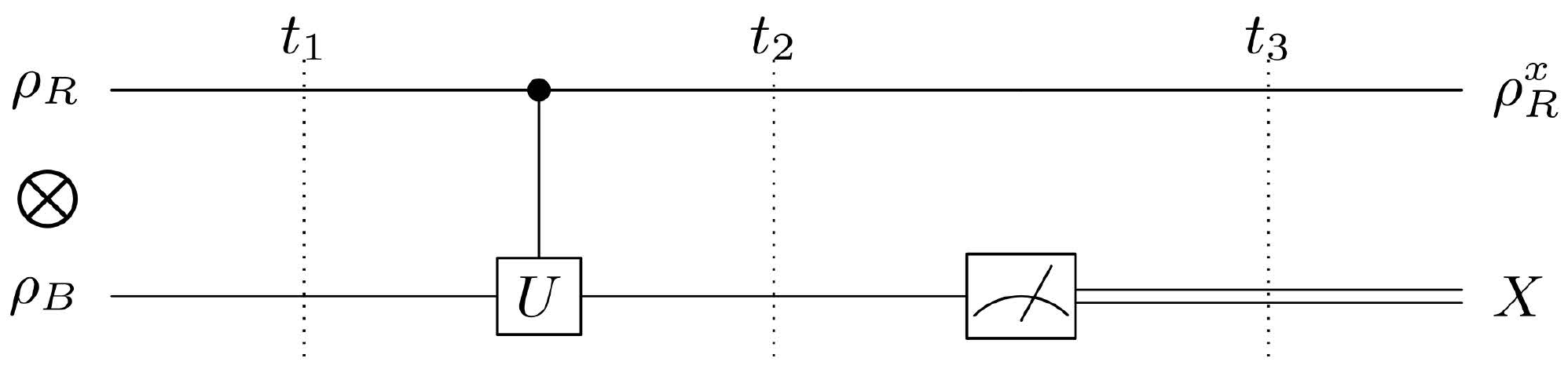}
\caption{\textbf{Uncertainty game as a quantum circuit.}
Initially, at time $t_1$, Alice's register $R$ and Bob's system $B$ do not share any correlations.
Then Alice makes a choice of the measured observable based on the state of the (possibly quantum) coin in $R$ by performing a conditional rotation $U$ on $B$. She then performs a measurement of the observable $S$ on $B$ to obtain the measurement outcome $X$. If the register $R$ is classical, i.e. it is diagonal on the standard basis, then these two operations of Alice effectively perform a random measurement of $S$ or $T$. If there is some non-zero coherence in register $R$, then the effective measurement can no longer be described as a random choice of one of the two observables. After that at time $t_3$ Alice sends $R$ to Bob. Bob then wants to guess Alice's outcome $X = x$ by trying to distinguish the states $\{\rho_R^x\}$. Note that if $R$ is classical, then the correlations between the two systems at time $t_2$ can also only be classical and all the states $\{\rho_R^x\}$ will be classical as well, implying that the optimal measurement of Bob corresponds to simply checking which one of the two observables Alice has chosen to measure. If $R$ contains coherence, then quantum correlations between the two registers can arise at time $t_2$ and Bob can better distinguish the states $\{\rho_R^x\}$ by performing a measurement that takes this coherence into account. Figure taken from~\cite{rozpkedek2017} with modifications under the licenses/CC BY 3.0 https://creativecommons.org/licenses/by/3.0/.
}
\label{fig:circuit}
\end{center}
\end{figure}

For the purpose of this paper it will be helpful to represent this game in a form of a quantum circuit as shown in FIG.~\ref{fig:circuit}. In this case let us assume that the measurement performed on register $B$ in this circuit corresponds to measuring observable $S$. Moreover, let us assume that the observable $T$ is related to $S$ through the relation $T = U^\dag S U$, where $U$ is the unitary operation shown on the circuit. Hence, if the classical coin contained in register $R$ is in state $\ket{0}$, then Alice measures observable $S$ on register $B$, while if the coin is in state $\ket{1}$, then Alice applies operation $U$ to the state on $B$, followed by the same measurement, which effectively leads to the measurement of the observable $T$ on $B$. After that, Bob measures the state on $R$ in the standard basis to find out what the outcome of the coin flip was and hence which observable has been chosen by Alice.

A complete mathematical description of this game, in which initially Bob does not know the outcome of the coin flip in $R$ requires us to set $\rho_R$ to a maximally mixed state. Then, Alice's measurement outcome $X=x$ leaves the register $R$ in the state $\rho_R^x$, and Bob's probability of guessing Alice's outcome is exactly the probability of how well he can distinguish all the states $\{\rho_R^x\}$. However, $R$ describes a random coin flip and therefore all $\{\rho_R^x\}$ will be diagonal in the standard basis (see Appendix~\ref{sec:guessGameAppendix} for details). This implies that Bob's optimal measurement is the $Z$-basis measurement which simply checks which one of the two observables Alice has measured, as discussed before.

Clearly, the classical coin flip used to choose the measurement of one of the two observables $S$ and $T$ inputs a classical randomness in the game and hence could be responsible for the arising inability of Bob to perfectly predict the measurement outcome of Alice, as suggested and then further investigated in~\cite{rozpkedek2017}. In this work, the authors analyze the consequences of removing this source of classical randomness by giving Bob access to the purification of that coin flip. In this way Bob has all the information about the corresponding choice of the observable to be measured and consequently this choice is now done on the quantum level. Clearly it is also possible that only some part of the purification of the coin flip is accessible to Bob and this is illustrated by the entangled registers $R$ and $P$ in FIG.~\ref{fig:game}, where $P$ is the register to which Bob never has access.

From the perspective of the quantum circuit in FIG.~\ref{fig:circuit}, for the generalized game the state on $R$ is no longer diagonal in the standard basis and so the coherence of $\rho_R$ implies that the choice of the measured observable is now performed through a quantum control. Moreover, after Alice performs her measurement, the resulting states $\{\rho_R^x\}$ are, in general, also no longer diagonal in the standard basis. Hence, Bob can now increase his guessing probability by applying a judiciously chosen measurement which extracts additional useful information from the off-diagonal coherence terms in $R$.

This guessing game enables us to seek a deeper understanding of the quantum uncertainty and to distinguish between the uncertainty stemming from Bob's lack of information (including the classical and the quantum information) and the intrinsic (unavoidable) uncertainty. We provide a high-level mathematical description of this guessing game framework in Appendix~\ref{sec:guessGameAppendix} while further details can be found in~\cite{rozpkedek2017}.

Let us first shed some light on the form of the state in register $R$. The form of this register determines the information that Bob has about the choice of the observable to be measured and therefore it determines his level of lack of knowledge about the measurement process. In the case of full lack of knowledge the two-dimensional register $R$ represents a random coin and so $\rho_R = \id/2$. In the case when Bob possesses all the information about the measurement process, $\rho_R$ would be a pure state and since we would like it to correspond to the scenario in which both measurements were chosen with equal probability, it is natural to set $\rho_R = \dyad{+}$, where $\ket{+} = \frac{1}{\sqrt{2}}\left(\ket{0} + \ket{1}\right)$.
One can then interpolate between the two cases by parameterizing $\rho_R$ using a $\gamma \in [0,1]$ parameter as follows:
\begin{equation}
\rho_R (\gamma)=\frac{1}{2}\left(\dyad{0}+\dyad{1}+\gamma \dyad{0}{1}+\gamma \dyad{1}{0}\right).
\label{eq:rhoRgammaMain}
\end{equation}
The physical meaning of $\gamma$ is discussed in Appendix~\ref{sec:guessGameAppendix}, while further details can be found in~\cite{rozpkedek2017}.

We note that we effectively have a whole family of guessing games, each of them corresponding to a specific configuration of the parameter set $(\gamma, d)$. Here $\gamma \in [0,1]$ is the coherence parameter described above, while $d = \{2,3,...\}$ describes the dimension of the game. Specifically, $d$ determines the number of possible outcomes of Alice's measurement and the dimension of the input state $\rho_B$.

In order to extract all the possible potential intrinsic uncertainty, the two measurements $S$ and $T$ that Alice performs are set to correspond to measuring in mutually unbiased bases. A natural choice for such bases is to set $S$ to be an observable corresponding to the measurement in the standard basis and $T$ to be an observable corresponding to the measurement in the Fourier basis.

Let us first have a quick look at the $d=2$ game. In this case the two measurements $S$ and $T$ correspond to the measurements in the standard and the Hadamard bases, respectively. After optimizing over all input states of Bob and his later measurement of the register $R$, it has been shown in~\cite{rozpkedek2017} that the maximum achievable guessing probability is given by:
\begin{equation}
P_{\text{guess}}^{\text{max}}(\gamma,d=2)=\frac{1}{2}\left(1+\frac{\sqrt{2+2\gamma^2}}{2}\right).
\end{equation}
In particular, $P_{\text{guess}}^{\text{max}}(\gamma,d=2)=1$ when $\gamma=1$. In this case, Bob can perfectly predict Alice's measurement outcome, and all the uncertainty is due to the lack of information. The work of~\cite{rozpkedek2017} also examines the link between uncertainty and the lack of information for higher-dimensional games with $d>2$. In these cases perfect guessing turns out to be no longer possible which shows the existence of the intrinsic uncertainty in those higher dimensions.

In the following, we implement the $d=2$ and $d=3$ guessing games, and experimentally study the relation between the coherence of the register $R$ and Bob's uncertainty about Alice's measurement outcome in order to verify the theoretical predictions of \cite{rozpkedek2017}. Specifically, for both the $d=2$ and $d=3$ guessing games with the chosen values of $\gamma > 0$, we observe a guessing probability which is larger than $P_{\text{guess}}^{\text{max}}(\gamma=0,d)$. In this way we verify that Bob's uncertainty arising in the scenario when the system $R$ is a classical coin, can be reduced by providing him with access to the purification of that classical coin flip. For the $d=2$ game we also observe that the larger the coherence parameter $\gamma$, the larger the experimentally observed guessing probability of Bob. Hence we can experimentally outperform the minimum possible amount of uncertainty for a given amount of revealed quantum information, by giving the guessing party additional quantum information about the state of the measurement apparatus. Finally, for the $d=2$ game with the largest possible value of $\gamma$ that we have been able to realize experimentally, the observed guessing probability becomes close to one. In other words, for the scenario in which we give the guessing party access to almost all the discussed quantum information, we observe almost no uncertainty at all which verifies the theoretical prediction of~\cite{rozpkedek2017}, that for the $d=2$ game there is no intrinsic uncertainty. The small amount of uncertainty that remains is directly established to be a result of the specific noise processes in our physical setup.

\begin{figure*}

\begin{center}
    	\includegraphics[width=1.8\columnwidth]{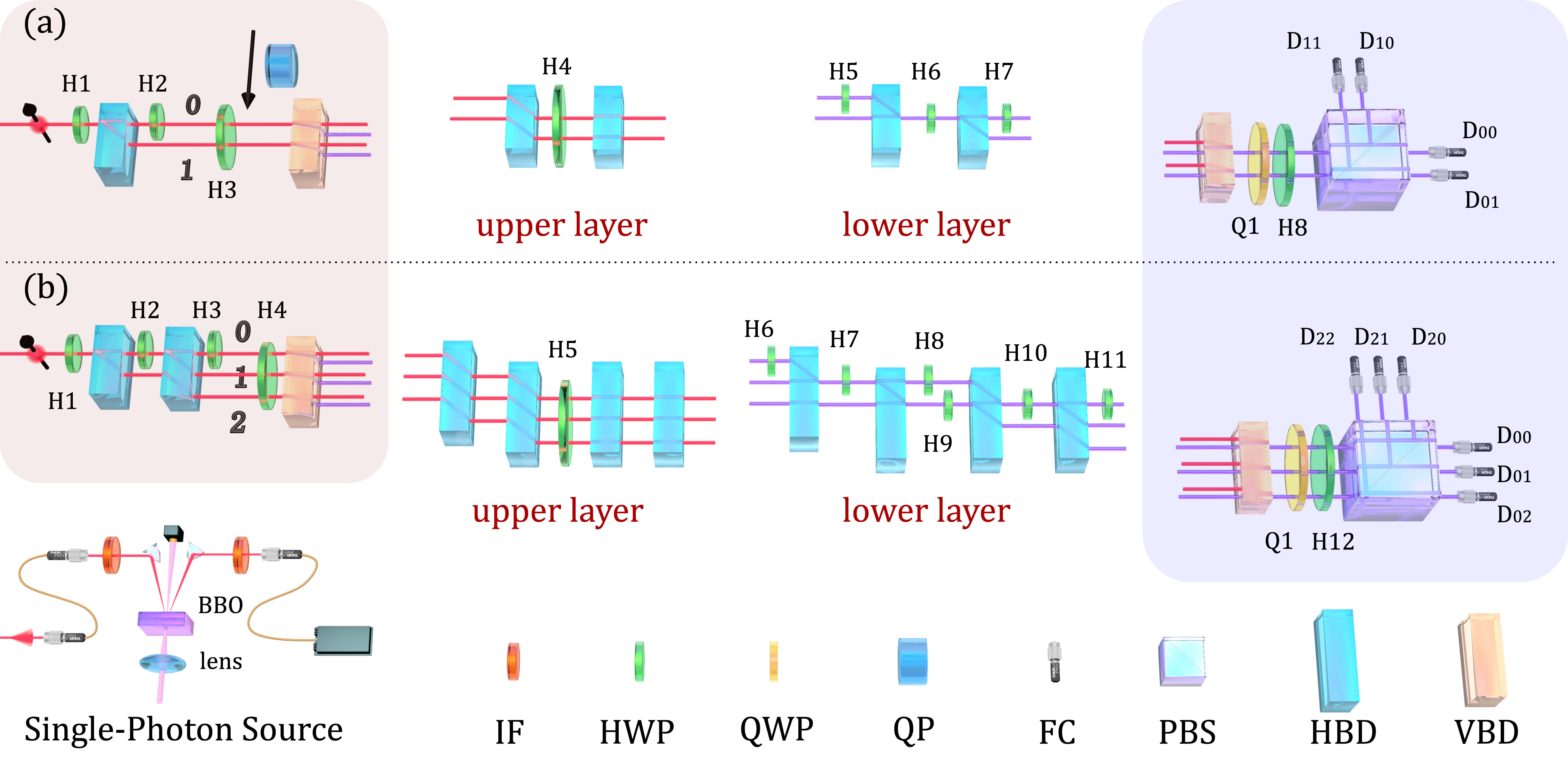}
        \caption{{\bf Experimental setup. (a) $d=2$ guessing game. (b) $d=3$ guessing game.} The single-photon is prepared by detecting one of the photons from a photon-pair generated in the Type-II spontaneous parametric down-conversion process. The whole setup consists of three modules: the state preparation part (red region), the controlled Fourier gate (white region), and the measurement part (purple region).  Firstly, Bob prepares the system $B$ in state $\rho_B$ and Alice prepares the register $R$ in state $\rho_R$, and those two systems are uncorrelated at module 1. Then a controlled Fourier gate is applied to the systems to correlate them.
        At last, Alice measures the system $B$ to obtain outcome $X$ and Bob measures the system $R$ in some optimal basis to help him guess $X$. In our experiment, the systems $B$ and $R$ are encoded in different degrees of freedom of a photon: the horizontally spatial modes marked as
        "0", "1" and "2" and the different path layers marked as upper layer "$u$" and lower layer "$l$", respectively. Therefore, if the register $R$ is in state $\ket{u}$ ($\ket{l}$), then the photon passes through the upper (lower) layer and undergoes an identity (Fourier) transformation as shown by the red (purple) lines. In the end, Alice needs to perform a non-demolition measurement, which is very difficult to realize in practice \cite{NDM}, before sending the system $R$ to Bob. Here we perform both measurements simultaneously to ensure the efficiency. Abbreviations: IF, interference filter; HWP, half-wave plate; QWP, quarter-wave plate; QP, quartz plate; FC, fiber coupler; PBS, polarizing beam splitter; HBD, horizontally placed beam displacer; VBD, vertically placed beam displacer; BBO, beta-barium-borate crystal. }\label{fig:setup2}
    \end{center}
\end{figure*}

In our experiment, we use the single photon system to implement the guessing game, and the basic idea is to use two independent DoFs of the photon to encode the system state $\rho_B$ and the register state $\rho_R$, respectively.
Specifically, as illustrated in FIG.~\ref{fig:setup2}, the system $B$ is encoded in the horizontal paths marked as "0", "1" and "2". The measurement basis choice register $R$ is encoded in the independent sets of paths marked as upper layer "$u$" and lower layer "$l$". More detailed information about the experimental implementation of guessing games can be found in Methods section.

\noindent\textbf{Results for the two-dimensional guessing game}

\begin{figure}
\begin{center}
\includegraphics[width=0.9\columnwidth]{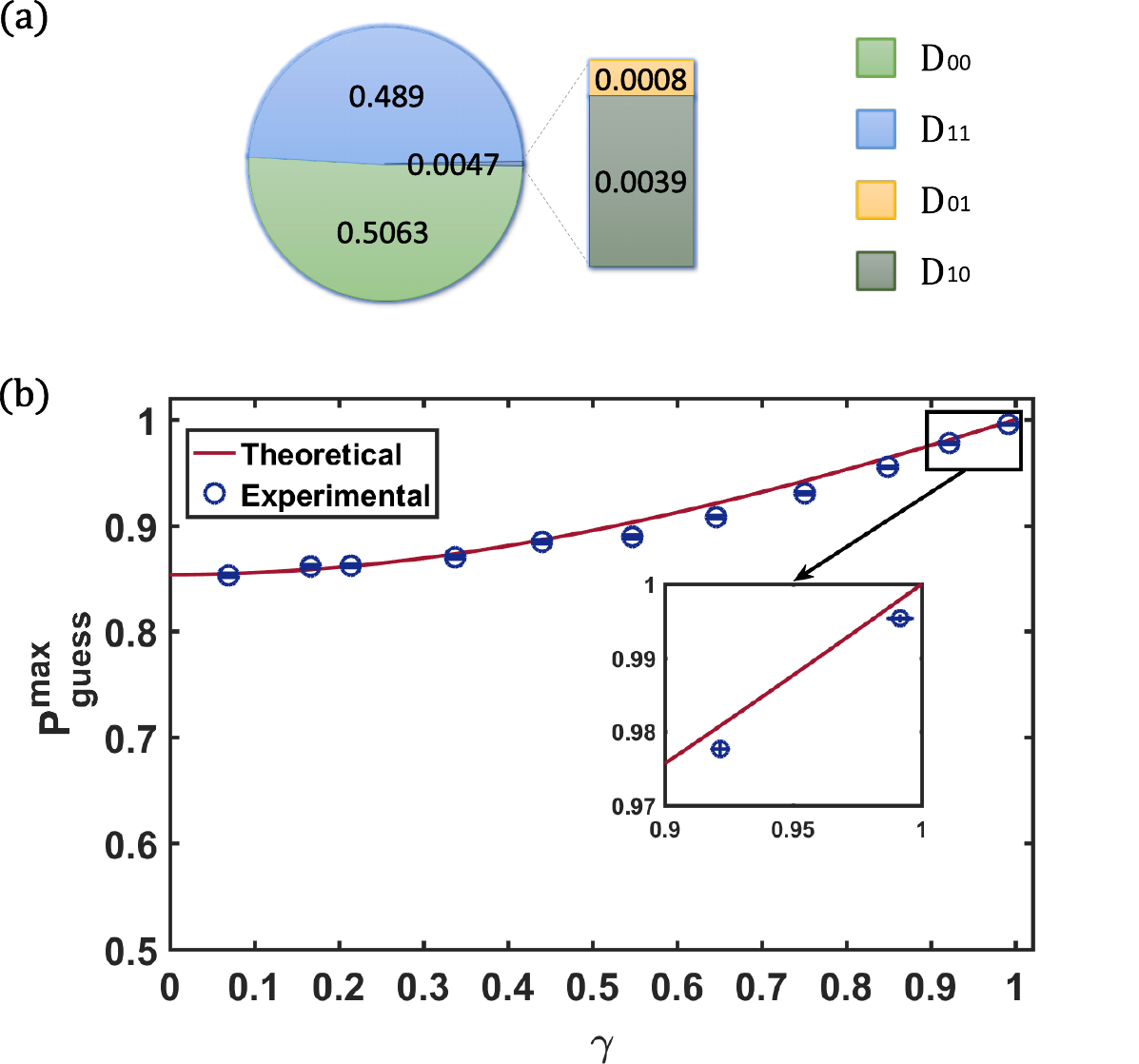}
        \caption{{\bf The experimental results for the $d=2$ guessing game.}
        From the experimentally prepared state $\rho_R^{\text{exp}}$ with the maximal purity, we estimate $\gamma=0.9918\pm0.0009$. With this $\gamma$, we obtain the maximal guessing probability $P_{\text{guess}}^{\text{exp}}(\gamma, d=2) =0.9953\pm0.0003$ and the detected probabilities for each output port are shown in FIG. (a). In FIG. (b), we vary the degree of coherence of the register state $\rho_R$ to find the relation between $P_{\text{guess}}^{\text{max}}$ and  $\gamma$. The analytical solution is plotted as the red line, while the experimental results are given as the blue circles. The $x$-bars are the standard deviations obtained by repeating the quantum state reconstruction algorithm for input data randomly generated from the experimentally obtained probability distributions. The $y$-bars are obtained directly from the detection probabilities in $D_{00}$, $D_{01}$, $D_{10}$ and $D_{11}$. }\label{fig:result1}
\end{center}
\end{figure}

While the classical randomness is adopted in the guessing game, Bob's maximum achievable guessing probability is $P_{\text{guess}}^{\text{max}}(\gamma=0,d=2)=(2+\sqrt{2})/4$. In our experiment, however, we observe that for 10 out of 11 data points with $\gamma>0$, $P_{\text{guess}}^{\text{exp}}(\gamma>0,d=2) > P_{\text{guess}}^{\text{max}}(\gamma=0,d=2)$. Here the superscript "exp" refers to the experimentally observed value, see the blue data points in FIG.~\ref{fig:result1}(b). This can be ascribed to the quantum information held in register $R$ and verifies that indeed there is uncertainty in the $\gamma=0$ game which comes from lack of information about the state of the purification register $P$.

Moreover, we see that $P_{\text{guess}}^{\text{exp}}$ increases with $\gamma$. Specifically for all $0 \leq \gamma < 0.9810$, we have observed an experimental value $P_{\text{guess}}^{\text{exp}}(\gamma + \delta,d=2)$ for some $0 < \delta< 0.2258$ such that $P_{\text{guess}}^{\text{exp}}(\gamma + \delta,d=2) > P_{\text{guess}}^{\text{max}}(\gamma,d=2)$, see
Appendix~\ref{sec:settings}, where we give the detailed values of $P_{\text{guess}}^{\text{exp}}$ and $P_{\text{guess}}^{\text{max}}$ for each $\gamma$. As $P_{\text{guess}}^{\text{max}}(\gamma,d=2)$, plotted as the solid red line in FIG.~\ref{fig:result1}(b), is the optimal guessing probability for a given $\gamma$, it is in fact an upper bound on that achievable probability. Hence, we have experimentally verified that for every $\gamma$ in that region we can perform better than the corresponding upper bound by giving Bob more access to the purification register (i.e. by experimentally increasing $\gamma$ to $\gamma + \delta$). Therefore our experiment verifies that indeed the more quantum information about the measuring process is given to Bob, the higher is the probability of him winning the game.

As we mentioned earlier, the optimal guessing probability for $\gamma=1$ is $P_{\text{guess}}^{\text{max}}(\gamma=1,d=2)=1$, which means that Bob can guess Alice's measurement result perfectly if he knows all the information of her measurement basis choice on the quantum level. In our experiment the highest value we observe is $P_{\text{guess}}^{\text{exp}}(\gamma,d=2)=0.9953\pm0.0003$, see FIG.~\ref{fig:result1}(a) where we show the detected probabilities for all the output ports for this scenario.
The fact that we cannot reach $P_{\text{guess}}^{\text{max}}(\gamma=1,d=2)=1$ can be ascribed to two main reasons. The first one is related to the fact that we cannot prepare the perfect state $\rho_R(\gamma = 1)$. Specifically, the maximal estimated $\gamma$ we obtained in the experiment is $\gamma=0.9918\pm0.0009$, and the fidelity between the experimentally prepared state and the theoretical state $\rho_R(\gamma=0.9918)$ is $0.9996$. The second reason is the fact that the visibility of the interferometer composed of the two vertically placed beam displacers (VBDs) stays about $0.99$ when collecting the data. This results in a dephasing error on the states $\rho_R^x(\gamma=0.9918, d=2, \rho_B)$. The detailed error analysis for $d=2$ guessing game is given in Appendix~\ref{sec:errord2}.

\begin{figure}
\begin{center}
    	\includegraphics[width=0.9\columnwidth]{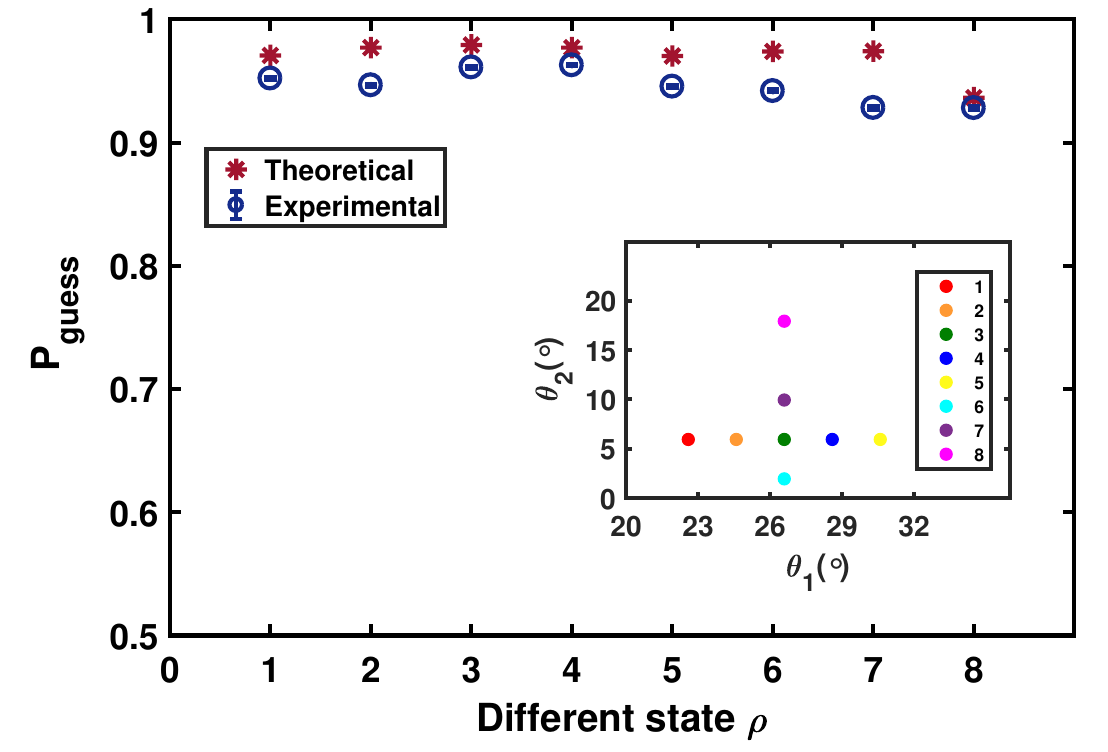}
        \caption{{\bf Different strategies for the $d=3$ guessing game.}
        The probabilities of the successful guessing in the experiment are shown as the blue circles with the theoretically predicted values shown as the red stars.
        For each strategy, the probe state is prepared with H1, H2, and two HBDs in FIG.~\ref{fig:setup2}(b). The best known strategy corresponds to data "3" and the corresponding setting of H1 ($\theta_1$) and H2 ($\theta_2$) for the input state preparation is shown as the green dot in the inset figure, meanwhile, the settings for other strategies are shown as the other colored dots. Notice that for the data point "8" the theoretically predicted value is much lower than for the other states. This is because the input state of data point "8" lies much further away from the best known strategy of data point "3" than all the other considered states, as can be seen on the inset.
        More information about the settings of the waveplates H1, H2, Q1, H12 and the detailed numerical values of the corresponding guessing probabilities are given in Appendix~\ref{sec:settings}. }\label{fig:result2}
    \end{center}
\end{figure}

\noindent\textbf{Results for the three-dimensional guessing game}
For the $d=3$ scenario, implementing the game for the largest $\gamma$ achievable in our experimental setup, given by $\gamma=0.9918$, and using the best known strategy results in the experimental guessing probability of $P_{\text{guess}}^{\text{exp}}(\gamma=0.9918,d=3) = 0.9611\pm0.001$ (see data "3" in FIG. \ref{fig:result2}). However, experimental procedures are subject to noise, which in many practical scenarios is non-isotropic and hence has a more severe effect on some states than others. Therefore it is possible that for our experimental setup the highest observed guessing probability could occur for a slightly different strategy than the one predicted in a noiseless scenario. To maximize our observed guessing probability and to obtain further insight into the effect of noise in our experiment, we test some other guessing strategies. Specifically, we choose various input states around the one stated above and modulate Bob's measurement to make sure the measurement is optimal for each state. From the results in FIG.~\ref{fig:result2}, we see that the highest successful guessing probability $P_{\text{guess}}^{\text{exp}}=0.9628\pm0.0009$ is achieved at data point "4", for which the input state is very close to the best probe state we found in theory.
Moreover, we note that compared with other data points, data points "6" and "7" have larger gaps to the theoretical values. That is mainly because the rotation of the wave-plates H2 introduces an unknown random phase in the interferometers. This issue is discussed in more detail in Appendix~\ref{sec:errord3}.

Similarly as in the $d=2$ game, we also observe in this case that the achieved $P_{\text{guess}}^{\text{exp}}(\gamma=0.9918,d=3) > P_{\text{guess}}^{\text{max}}(\gamma=0,d=3)=\frac{1}{2}\left(1+\frac{1}{\sqrt{3}}\right)$. Hence we have experimentally demonstrated that lack of information is also a significant source of uncertainty in the $d=3$ game. Comparing our experimentally observed value of the guessing probability for $\gamma=0.9918$ with the highest known achievable guessing probability in the noiseless scenario using the strategy from~\cite{rozpkedek2017}, we see that our result also outperforms those scenarios for the values of $\gamma$ up to more than 0.9. Unfortunately the optimal strategy for $d=3$ game with $\gamma >0$ is not known, and therefore we cannot claim that we outperformed the optimal strategies for all those lower values of $\gamma$.
However, our achieved high guessing probability gives a strong experimental indication that also in the $d=3$ game giving Bob access to more quantum information about the purifying register $P$, enables him to win with higher probability.

On the other hand, our results also provide an insight into the existence of the intrinsic uncertainty in the $d=3$ game. As the theoretical analysis in~\cite{rozpkedek2017} has shown it is not possible to achieve perfect guessing for that game. This is unlike in the $d=2$ case, where all the uncertainty can be contributed to the lack of information. The highest known achievable guessing probability for the $d=3$ game in the noiseless scenario is $P_{\text{guess}}(\gamma=1, d=3)=0.9793$. Let us now compare our experimentally observed values with this theoretical prediction. We will focus here on the data point "3" as our experimental setup was optimized for this setting thus making the error analysis easier for this data point, while the increase in the observed guessing probability for data point "4" is small. Comparing with the best known $P_{\text{guess}}(\gamma=1, d=3)$ for the noiseless case, the guessing probability we achieved in the experiment for data point "3" has an error gap of $p_{\text{gap}} = 0.0182$ to this theoretical value, which can be ascribed to two aspects. On the one hand, in our experiment we use $\gamma=0.9918$ instead of $\gamma=1$; on the other hand, there are experimental errors. In Appendix~\ref{sec:errord3} we verify that the observed error gap is consistent with our error model based on the characterized components of the setup. In particular, the experimental errors correspond to the state preparation errors and the dephasing errors inside the interferometers in the setup. Having verified the origin of this error gap, which we can refer to as a gap due to lack of information, we note that in constitutes only a smaller part of the total observed uncertainty gap $1 - P_{\text{guess}}^{\text{exp}}(\gamma=0.9918,d=3)$ for data point "3". In particular:
\begin{equation}
\frac{p_{\text{gap}}}{1 - P_{\text{guess}}^{\text{exp}}(\gamma=0.9918,d=3)} = 0.4679.
\end{equation}
This shows that if the best known theoretical strategy was indeed the optimal one, then more than half of the total experimentally observed uncertainty gap would not come from lack of information but from the intrinsic uncertainty. This observation gives an experimental support to the claim that intrinsic uncertainty is present in the $d=3$ game.

\section*{Discussion}
\label{sec:conclusion}

Our work experimentally studies the entropic formulation of quantum uncertainty within the guessing game framework. We experimentally verify that lack of quantum information about the register governing the choice of the measured observable is a key contributor to the arising uncertainty. Our results have been obtained by experimentally implementing a $d=2$ and $d=3$ guessing games. We also see, especially for the $d=2$ game, that the more quantum information about the measurement process can be accessed by Bob, the higher his chance of winning the game. We also observed guessing probability of almost one for the case when almost all the information about the measurement process was made available to Bob, confirming the result of~\cite{rozpkedek2017} that for the $d=2$ game there is no intrinsic uncertainty. Finally, the obtained data for the $d=3$ game supports the result of~\cite{rozpkedek2017} that there exists intrinsic uncertainty for the $d=3$ game.

These results have implications for various cryptographic protocols that make use of measurements in mutually unbiased bases. In particular for protocols that perform measurements in BB84 bases~\cite{Bennett_84}, we see that it is vital for the purification of the coin determining the measurement basis, to be inaccessible to the eavesdropper. Otherwise the security may be compromised, and in the case when the eavesdropper could later have access to the entire purification of the coin, they could be able to always guess the measurement outcome and hence e.g. obtain the entire key in BB84 QKD~\cite{Bennett_84,scarani2009security}.

Moreover, our work forms an important step in the experimental development of quantum optical technologies based on multidimensional systems.
The development of our setup contributes to the existing linear optics toolbox through the realization of the controlled three-dimensional quantum Fourier transform.
Here, the method we use to implement the three-dimensional quantum Fourier transform can be generalized to arbitrary unitary transformations by regulating the settings of the waveplates. When extending to a much higher dimension, one of the obstacles lies in the relatively large volume of the calcite beam displacer, which must enable multiple beams to pass through simultaneously. For instance, the sizes of the beam displacers in our experiment are approximately 8~mm$\times$15~mm$\times$37.71~mm. An efficient way to overcome the size problem is by stacking a series of PBSs, just like in \cite{pan2018,guo2020,pan2020,pan2021}. Another problem that one needs to consider is phase stability. As the complexity of the setup increases, an active phase stabilization system may need to be built.

Furthermore, our setup also offers the possibility to further investigate the wave-particle duality~\cite{WPD1, WPD2, WPD3, WPD4} and its connection to the uncertainty principle~\cite{WPD5}. Finally, we note that a further refinement of the controlled Fourier transform to the case in which the control system is also a qutrit and the target system undergoes a transformation to one of the three incompatible measurements would enable us to investigate experimentally the recent results of~\cite{plesch2018loss,doda2020choice}. In these works the authors extend the game of~\cite{rozpkedek2017} to measuring more than two observables. Interestingly, they show that for the game in which $B$ is two-dimensional, guessing probability of one can be achieved independently of how many measurements are considered. However, if $B$ is more than two-dimensional and more than two measurements are considered, then they show that whether perfect guessing is possible depends on the specific choice of the incompatible measurements. These extensions of the original game for the scenario with three measurements could potentially be implemented on the modification of our setup.
\section*{Methods}
\noindent\textbf{Single-photon source}

In both the $d=2$ guessing game (FIG. \ref{fig:setup2}(a)) and $d=3$ guessing game (FIG. \ref{fig:setup2}(b)), pairs of photons of $808$~nm are generated by the spontaneous parametric down-conversion (SPDC) process with a $100$~mW, $404$~nm single-frequency laser ($<5$~MHz Linewidth) pumping a type-II BBO (beta-barium-borate) crystal. Then one of the photons is fed to the experimental setup as the signal photon, which is heralded by the detection of the other photon from the pair.

\noindent\textbf{Experimental implementation of guessing games}

The system state $\rho_B$ is prepared with the HWPs, (specifically H1 in FIG.~\ref{fig:setup2}(a), H1 and H2 in FIG.~\ref{fig:setup2}(b)) and HBDs, which sort the input beam into the horizontally parallel beams with different polarized directions \textit{H} and \textit{V} (\textit{H}, horizontally polarized direction; \textit{V}, vertically polarized direction).
A $45^{\circ}$ oriented HWP (H2 in FIG.~\ref{fig:setup2}(a) and H3 in FIG.~\ref{fig:setup2}(b)) is inserted in path "0" to unify the photon's polarization directions in different paths. Then a $22.5^{\circ}$ HWP prepares the polarization of the photon in all paths in a state $1/\sqrt{2}(\ket{H}+\ket{V})$ (H3 in FIG.~\ref{fig:setup2}(a) and H4 in FIG.~\ref{fig:setup2}(b)). After that a VBD directs the \textit{H} photon to the upper layer $\ket{u}$ (red lines) and \textit{V} photon to the lower layer $\ket{l}$ (purple lines), hence preparing the control state $1/\sqrt{2}(\ket{u}+\ket{l})$ on the register $R$. Then, depending on whether the photon passes through the upper layer or lower layer, it will undergo either the $\id$ operation or the Fourier operation. In our experimental setup the parallel-path structure of the interference is stable, because all the light beams are affected by the environmental turbulences, such as temperature fluctuation and vibrations, in nearly the same way~\cite{o2003demonstration}. Then Bob uses the second VBD to convert the path DoF corresponding to the upper and lower layer into the polarization DoF and uses a quarter-wave plate (QWP, Q1), an HWP (H8 in FIG.~\ref{fig:setup2}(a) and H12 in FIG.~\ref{fig:setup2}(b)) and a polarization beam splitter (PBS) to distinguish the quantum states $\rho_R^x$ in order to guess Alice's measurement outcome $X$. We note that since both registers $R$ and $B$ are encoded in different DoF of the same photon, in the experiment a simultaneous measurement of both registers is performed at once. Specifically, the click in the output port $D_{ij}$ corresponds to Bob's guessing outcome $i$ for Alice's measurement outcome $j$. Therefore, Bob's goal is to set Q1 and H8 (H12) in such a way so that the probability of detection in the ports $D_{ii}$ is maximized.

For the $d=2$ game, one of the input states of Bob that is optimal for all $\gamma$ is the pure state $\ket{\psi}_B \propto \ket{0}+\ket{-}$, where $\ket{-}=1/\sqrt{2}(\ket{0}-\ket{1})$. This state is prepared by setting the orientation angle of H1 to $11.3^{\circ}$. Meanwhile, to observe the relation between $P_{\text{guess}}^{\text{max}}(\gamma,d=2)$ and $\gamma$, we place the quartz plate (QP) before the VBD to decrease the coherence between $\ket{u}$ and $\ket{l}$. Now the polarization of the photon is coupled by the QP to its frequency distribution realizing the dephasing channel, and the value of $\gamma$ is tuned by changing the thickness of the QP. Before the VBD, we perform the standard tomography process to reconstruct the experimentally generated register state $\rho_R^{\text{exp}}$. The value of $\gamma$ is estimated by approximating $\rho_R^{\text{exp}}$ by an ideal register state $\rho_R(\gamma)$ given in Eq.~\eqref{eq:rhoRgammaMain}. That is, $\gamma$ of $\rho_R^{\text{exp}}$ is taken to be the value of that parameter for this $\rho_R(\gamma)$ which has the highest fidelity to $\rho_R^{\text{exp}}$. We find that for each obtained $\gamma$ the fidelity between $\rho_R^{\text{exp}}$ and the corresponding $\rho_R(\gamma)$ is higher than $0.9995$. Finally, the guessing probability is obtained by summing the detection probabilities in output ports $D_{00}$ and $D_{11}$. More details about the thicknesses of quartz plates, the angles of Q1 and H8, as well as the detailed numerical values of the corresponding experimental results are provided in Appendix~\ref{sec:settings}.

For the $d=3$ game we focus on the single scenario corresponding to the largest possible $\gamma$ that we could achieve in our experiment.
We then investigate the optimal known strategy for that $\gamma$.
The best probe states for the $d=3$ game that we found, established using the procedure from~\cite{rozpkedek2017} have a nice property that for all $\gamma$ the optimal measurement for Bob aiming to distinguish the three possible qubit states $\rho_R^x$ is actually a projective measurement. This measurement aims to distinguish only two out of the three possible states, corresponding to the two dominant outcomes of Alice. Specifically, for the best known input state we consider, the dominant outcomes are 0 and 2. The corresponding projective measurement performed on the register $R$ has POVM elements $\{M_0, M_1 = \mathbf{0}, M_2\}$, where $M_0$ and $M_2$ are projectors. This explains why the first index of detectors $D$ in  FIG.~\ref{fig:setup2}(b) takes only the value 0 or 2.

In our experiment, the highest amount of coherence in the register $R$ which we achieved is $\gamma=0.9918$.
A corresponding best probe state we found for the $d=3$ game is the state $\ket{\psi}_B=a_1\ket{0}+a_2\ket{1}+a_3\ket{2}$ with the coefficients $a_1=0.0938 + 0.5786i$, $a_2=0.0109 - 0.1218i$ and $a_3=0.8009$. More detailed information about the probe states preparation, the optimal measurements, and the guessing probabilities we obtained are given in Appendix~\ref{sec:settings}.

\noindent\textbf{Three-dimensional Fourier gate}

We note that in the $d=3$ guessing game we implement the three-dimensional Fourier operation based on the idea of the scheme proposed in \cite{clements2016optimal}.
In the original scheme, the single-qubit rotation operator $R_y$ represents a variable beam splitter, which is realized by an interferometer built with two 50:50 beam splitters. The phase difference between the two arms of the interferometer is adjusted to change the ratio of the light beams in two output ports. In our work, we develop a HBD-HWP-HBD structure to realize the operator $R_y$, which uses much fewer elements compared with the method with 50:50 beam splitters. Hence our scheme is much more friendly to the experimental implementation. Owing to the introduction of the polarization-dependent beam splitter, HBD, which enables the transformation between the path DoF and the polarization DoF, the photon's paths can be efficiently manipulated by the polarization controller element HWP instead of the interferometer.

Let us now briefly discuss how we quantify the performance of this Fourier gate. After applying the ideal Fourier operation to the input state $\ket{w_j}=1/\sqrt{3}\sum_{k=0}^2w^{-jk}\ket{k}$, where $j=0,1,2$, $w=e^{2i\pi/3}$, we will obtain the corresponding output state $\ket{j}$, therefore the probability to detect a photon in output mode $i$ when inputting state $\ket{w_j}$ into our Fourier gate implementation should be $\delta_{ij}$. In our experiment, the average probability for detecting the photon in the right output mode is $0.9771\pm0.0006$, which can be obtained only when the Fourier operation works well. The detailed information about how to implement and estimate the quality of the Fourier operation are given in the Appendix~\ref{sec:FourierGate}. Moreover, we analyze the main factors limiting its performance by considering a three-dimensional dephasing model in Appendix~\ref{sec:errord3}.

\section*{Data availability}
All the data that support the results of the current work are available from the corresponding authors upon reasonable request.

\section*{Code availability}
The codes for simulation and data processing are available from the corresponding authors upon reasonable request.

\section*{Acknowledgments}

We would like to greatly thank Jan Ko\l{}ody\'nski for help with modelling dephasing noise in interferometers. We are also very grateful to J\k{e}drzej Kaniewski for valuable feedback on the manuscript. The work at the University of Science and Technology of China is supported by the National Natural Science Foundation of China (Grants No. 11804410, 11974335, 11574291, 11774334 and 61905234) and the China Postdoctoral Science Foundation (Grant No. 2020M682001).

\section*{Author contribution}

Y.Y.Z. and F.R. contributed equally to this work. Y.Y.Z. is the main experimental author and F.R. the theory author of this work. Y.Y.Z. designed and performed the experiment with the help from Z.H. and K.D.W., and F.R. solved the optimization problems for the optimal device settings. Y.Y.Z. and F.R. analyzed the data, constructed the error models, and wrote the manuscript. G.Y.X., C.F.L. and G.C.G. supervised the project.

\section*{COMPETING INTERESTS}

The authors declare no competing interests.

\bibliographystyle{nature}
\bibliography{manuscript}

\begin{thebibliography}{10}

\bibitem{1927}
Heisenberg, W.
\newblock In {\em Original Scientific Papers Wissenschaftliche
  Originalarbeiten},  478--504. Springer (1985).

\bibitem{kennard1927quantenmechanik}
Kennard, E.~H.
\newblock {\em Z Phys.}{ \bf 44}(4), 326--352 (1927).

\bibitem{robertson1929}
Robertson, H.~P.
\newblock {\em Phys. Rev.}{ \bf 34}, 163--164 Jul  (1929).

\bibitem{2017entropic}
Coles, P.~J., Berta, M., Tomamichel, M., and Wehner, S.
\newblock {\em Rev. Mod. Phys.}{ \bf 89}, 015002 Feb  (2017).

\bibitem{maassen1988}
Maassen, H. and Uffink, J. B.~M.
\newblock {\em Phys. Rev. Lett.}{ \bf 60}, 1103--1106 Mar  (1988).

\bibitem{berta2010}
Berta, M., Christandl, M., Colbeck, R., Renes, J.~M., and Renner, R.
\newblock {\em Nat. Phys.}{ \bf 6}(9), 659 (2010).

\bibitem{rozpkedek2017}
Rozp{\k{e}}dek, F., Kaniewski, J., Coles, P.~J., and Wehner, S.
\newblock {\em New J. Phys.}{ \bf 19}(2), 023038 (2017).

\bibitem{nielsen2011quantum}
{Nielsen}, M.~A. and {Chuang}, I.~L.
\newblock {\em Quantum Computation and Quantum Information: 10th Anniversary
  Edition}.
\newblock  (2011).

\bibitem{Shor1994}
Shor, P.
\newblock In {\em Proceedings 35th Annual Symposium on Foundations of Computer
  Science},  124--134,  (1994).

\bibitem{wootters1989}
Wootters, W.~K. and Fields, B.~D.
\newblock {\em Ann Phys}{ \bf 191}(2), 363--381 (1989).

\bibitem{miles2013}
Giovannini, D., Romero, J., Leach, J., Dudley, A., Forbes, A., and Padgett,
  M.~J.
\newblock {\em Phys. Rev. Lett.}{ \bf 110}, 143601 Apr  (2013).

\bibitem{zeilinger2006}
Grblacher, S., Jennewein, T., Vaziri, A., Weihs, G., and Zeilinger, A.
\newblock {\em New J. Phys.}{ \bf 8}(75) (2006).

\bibitem{FTsuperconducting}
Yurtalan, M.~A., Shi, J., Kononenko, M., Lupascu, A., and Ashhab, S.
\newblock {\em Phys. Rev. Lett.}{ \bf 125}, 180504 Oct  (2020).

\bibitem{FTtrappedions}
Klimov, A.~B., Guzm\'an, R., Retamal, J.~C., and Saavedra, C.
\newblock {\em Phys. Rev. A}{ \bf 67}, 062313 Jun  (2003).

\bibitem{FTAOM}
Brandt, F., Hiekkam\"{a}ki, M., Bouchard, F., Huber, M., and Fickler, R.
\newblock {\em Optica}{ \bf 7}(2), 98--107 Feb  (2020).

\bibitem{FTtimebin}
Lu, H.-H., Hu, Z., Alshaykh, M.~S., Moore, A.~J., Wang, Y., Imany, P., Weiner,
  A.~M., and Kais, S.
\newblock {\em Adv. Quantum Technol.}{ \bf 3}(2), 1900074.

\bibitem{FTNMR}
{Dogra}, S., {Arvind}, and {Dorai}, K.
\newblock {\em Phys. Lett. A}{ \bf 378}(46), 3452--3456 October  (2014).

\bibitem{clements2016optimal}
Clements, W.~R., Humphreys, P.~C., Metcalf, B.~J., Kolthammer, W.~S., and
  Walsmley, I.~A.
\newblock {\em Optica}{ \bf 3}(12), 1460--1465 (2016).

\bibitem{NDM}
Xia, K.
\newblock In {\em Photon Counting-Fundamentals and Applications}. InTech
  (2018).

\bibitem{Bennett_84}
Bennett, C.~H. and Brassard, G.
\newblock In {\em International Conference on Computer System and Signal
  Processing, IEEE, 1984},  175--179,  (1984).

\bibitem{scarani2009security}
Scarani, V., Bechmann-Pasquinucci, H., Cerf, N.~J., Du\ifmmode~\check{s}\else
  \v{s}\fi{}ek, M., L\"utkenhaus, N., and Peev, M.
\newblock {\em Rev. Mod. Phys.}{ \bf 81}, 1301--1350 Sep  (2009).

\bibitem{pan2018}
Wang, X.-L., Luo, Y.-H., Huang, H.-L., Chen, M.-C., Su, Z.-E., Liu, C., Chen,
  C., Li, W., Fang, Y.-Q., Jiang, X., Zhang, J., Li, L., Liu, N.-L., Lu, C.-Y.,
  and Pan, J.-W.
\newblock {\em Phys. Rev. Lett.}{ \bf 120}, 260502 Jun  (2018).

\bibitem{guo2020}
Hu, X.-M., Xing, W.-B., Liu, B.-H., Huang, Y.-F., Li, C.-F., Guo, G.-C., Erker,
  P., and Huber, M.
\newblock {\em Phys. Rev. Lett.}{ \bf 125}, 090503 Aug  (2020).

\bibitem{pan2020}
Zhong, H.-S., Wang, H., Deng, Y.-H., Chen, M.-C., Peng, L.-C., Luo, Y.-H., Qin,
  J., Wu, D., Ding, X., Hu, Y., Hu, P., Yang, X.-Y., Zhang, W.-J., Li, H., Li,
  Y., Jiang, X., Gan, L., Yang, G., You, L., Wang, Z., Li, L., Liu, N.-L., Lu,
  C.-Y., and Pan, J.-W.
\newblock {\em Science}{ \bf 370}(6523), 1460--1463 (2020).

\bibitem{pan2021}
Zhong, H.-S., Deng, Y.-H., Qin, J., Wang, H., Chen, M.-C., Peng, L.-C., Luo,
  Y.-H., Wu, D., Gong, S.-Q., Su, H., Hu, Y., Hu, P., Yang, X.-Y., Zhang,
  W.-J., Li, H., Li, Y., Jiang, X., Gan, L., Yang, G., You, L., Wang, Z., Li,
  L., Liu, N.-L., Renema, J.~J., Lu, C.-Y., and Pan, J.-W.
\newblock {\em Phys. Rev. Lett.}{ \bf 127}, 180502 Oct  (2021).

\bibitem{WPD1}
Tang, J.-S., Li, Y.-L., Xu, X.-Y., Xiang, G.-Y., Li, C.-F., and Guo, G.-C.
\newblock {\em Nat. Photonics}{ \bf 6}(9), 600 (2012).

\bibitem{WPD2}
Peruzzo, A., Shadbolt, P., Brunner, N., Popescu, S., and O'Brien, J.~L.
\newblock {\em Science}{ \bf 338}(6107), 634--637 (2012).

\bibitem{WPD3}
Kaiser, F., Coudreau, T., Milman, P., Ostrowsky, D.~B., and Tanzilli, S.
\newblock {\em Science}{ \bf 338}(6107), 637--640 (2012).

\bibitem{WPD4}
Ionicioiu, R. and Terno, D.~R.
\newblock {\em Phys. Rev. Lett.}{ \bf 107}, 230406 Dec  (2011).

\bibitem{WPD5}
Coles, P.~J., Kaniewski, J., and Wehner, S.
\newblock {\em Nat. Commun.}{ \bf 5}, 5814 (2014).

\bibitem{plesch2018loss}
Plesch, M. and Pivoluska, M.
\newblock {\em New J. Phys.}{ \bf 20}(2), 023018 (2018).

\bibitem{doda2020choice}
Doda, M., Pivoluska, M., and Plesch, M.
\newblock {\em Phys. Rev. A}{ \bf 103}, 032206 Mar  (2021).

\bibitem{o2003demonstration}
O'Brien, J.~L., Pryde, G.~J., White, A.~G., Ralph, T.~C., and Branning, D.
\newblock {\em Nature}{ \bf 426}(6964), 264 (2003).

\bibitem{helstrom1969quantum}
Helstrom, C.~W.
\newblock {\em J Stat Phys}{ \bf 1}(2), 231--252 (1969).

\bibitem{reck1994}
Reck, M., Zeilinger, A., Bernstein, H.~J., and Bertani, P.
\newblock {\em Phys. Rev. Lett.}{ \bf 73}, 58--61 Jul  (1994).

\bibitem{demkowicz2015}
Demkowicz-Dobrza{\'n}ski, R., Jarzyna, M., and Ko{\l}ody{\'n}ski, J.
\newblock {\em Prog. Opt.}{ \bf 60}, 345--435 (2015).

\end{thebibliography}

\newpage
\hspace{10cm}
\newpage
\appendix
\onecolumngrid

\section{The guessing game}\label{sec:guessGameAppendix}

\subsection{Quantifying lack of information}
In the general guessing game considered, the state of the register $R$ is given by
 \begin{equation}
 \rho_R=\frac{1}{2}(\dyad{0}+\dyad{1}+\gamma(\dyad{0}{1}+\dyad{1}{0})),
 \label{eq:rhoR}
 \end{equation}
 where $\gamma \in [0,1]$, and states $\ket{0}$ and $\ket{1}$ of $R$ are associated with the measurement of $S$ and $T$ respectively. We note that we assume that any possible complex phase in $\rho_R$ is also known to Bob and therefore $\gamma$ can be restricted to a real-valued parameter as shown in~\cite{rozpkedek2017}.

 To better understand the meaning of $\gamma$, which determines the coherence of $\rho_R$, and its relation to Bob's lack of information about the system $P$ which purifies $R$, let us recall how we defined those systems. Here we will follow the definitions and interpretations presented in~\cite{rozpkedek2017}. Specifically, even though Bob is given access to $R$, we emphasize that he does not have access to $P$ in our guessing game. Hence, we can think of $P$ as representing Bob's lack of information.

For example, for the classical game in which Bob sees the choice of the measurement basis as a random coin flip, $\rho_R = \id/2$. Then the purification of $R$ is a maximally entangled state such as
\begin{align}
\ket{\xi_{RP}} = \frac{1}{\sqrt{2}}\left(\ket{0}_R \ket{0}_P + \ket{1}_R \ket{1}_P\right)\, .
\end{align}
If $\rho_R$ is pure, then $P$ is in a tensor product with $R$ i.e.,
\begin{align}
\ket{\xi_{RP}}=\ket{\xi_R} \otimes \ket{\xi_P}.
\end{align}
Since in the classical game both $S$ and $T$ are measured with equal probability, a natural extension when the purification of the coin is included in $R$ is to set $\ket{\xi_R} = \frac{1}{\sqrt{2}}\left(\ket{0} + \ket{1}\right)$. Clearly the case when the initial state is maximally entangled, corresponds to $\rho_R = \id/2$ and so in this case $P$, to which Bob does not have access, holds the maximal amount of information useful to Bob. Of course if $R$ is already pure then P does not contain any additional information that Bob could use.
\\ \indent
Here we recall how~\cite{rozpkedek2017} interpolates between these two extremes. Let $C$ denote a classical coin. Then clearly $C$ must be part of R. However, additionally $R$ and $P$ consist of many environmental subsystems $E_1,\ldots,E_n$, each of which holds a small amount of information that will be useful to Bob. Then Bob's lack of information can be quantified by the number of the environment systems that are part of $P$ instead of part of $R$.

That is $R = CE_1\ldots E_j$ and $P = E_{j+1}\ldots E_n$. In~\cite{rozpkedek2017} it is then shown that the continuous parameter $\gamma \in [0,1]$ can be used to quantify the number of environmental subsystems included in $R$ in the limit $n \rightarrow \infty$.

\subsection{Optimal guessing probability}
The full evolution of the quantum states on registers $B$ and  $R$ is provided in~\cite{rozpkedek2017}. Here we provide the key information that allows us to pose the optimization problem for finding the optimal guessing probability. After Alice's measurement, the quantum-classical state between the register $R$ and the outcome $X$ is expressed as
\begin{equation}
 \rho_{RX}(\gamma, d, \rho_B)=\sum_x\tilde{\rho}_R^x(\gamma, d, \rho_B)\otimes\dyad{x}_x,
 \end{equation}
where
\begin{equation}
\tilde{\rho}_R^x(\gamma, d, \rho_B)=\frac{1}{2}\left[
 \begin{matrix}
 \bra{x}\rho_B\ket{x}  & \gamma\bra{x}\rho_B F^{\dag}\ket{x} \\
   \gamma\bra{x}F\rho_B\ket{x} & \bra{x}F\rho_BF^{\dag}\ket{x} \\
  \end{matrix}
  \right]
 \end{equation}
is the sub-normalized post-measurement state of the register $R$. Let us denote the corresponding normalized state as $\rho_R^x(\gamma, d, \rho_B) = \tilde{\rho}_R^x(\gamma, d, \rho_B)/p_x(d,\rho_B)$, where $p_x(d,\rho_B) = \text{Tr}(\tilde{\rho}_R^x(\gamma, d, \rho_B))$. Bob then tries to guess the outcome $X=x$ after determining which state $\rho_R^x(\gamma, d, \rho_B)$ he has received. Now the guessing problem becomes a state discrimination problem.
Finally, the maximal guessing probability is achieved by optimizing the input state $\rho_B$ and the corresponding measurement on $R$:
\begin{equation}
P^{\text{max}}_{\text{guess}}(\gamma, d)=\max_{\rho_B}\max_{\{M_x\}}\sum_{x=0}^{d-1}p_x(d,\rho_B)\text{Tr}[M_x\rho_R^x(\gamma, d, \rho_B)].
\label{eq:Pguessmaxdef}
 \end{equation}

For $d=2$, Helstrom has found the optimal measurements and the corresponding maximum probability of correctly distinguishing between two quantum states analytically~\cite{helstrom1969quantum}. His result makes it possible to easily find the optimal input state of Bob and hence to analytically calculate $P_{\text{guess}}^{\text{max}}$ as shown in~\cite{rozpkedek2017}. For $d>2$ no analytical solution to the optimisation problem~\eqref{eq:Pguessmaxdef} is known due to its non-convex nature. Therefore for higher-dimensional games that involve distinguishing more than two states we use numerical techniques described in~\cite{rozpkedek2017} that unfortunately cannot guarantee the global optimality of the found solution. Nevertheless, an analytical argument described in~\cite{rozpkedek2017} shows that $P^{\text{max}}_{\text{guess}}(\gamma, d>2) < 1$.

\section{Device settings for the implementation of the guessing game}
\label{sec:settings}
In this section we provide the numerical values of the settings of the optical components in our experimental setup. All the components are referred to according to their labels in FIG.~\ref{fig:setup2}.

\subsection{Settings for the $d=2$ game}
\label{sec:settingsd2}
In the $d=2$ guessing game, the wave plate H1 is rotated by $\theta_1$ to prepare the optimal input state $\ket{\psi}_B$ in basis $\ket{H}$ and $\ket{V}$.
The HBD is used to encode the polarization state into the path DoF by displacing the \textit{V} component into path-0, and the \textit{H} component into path-1 with a 4-mm lateral displacement. Then H2 unifies the polarization of the photon in different paths, and the system state becomes
\begin{equation}
\ket{\psi}_B=\cos2\theta_1\ket{0}-\sin2\theta_1\ket{1}.
\end{equation}
Here $\theta_1$ is set to $11.3^{\circ}$ to prepare the optimal input state.

\begin{figure}
    	\includegraphics{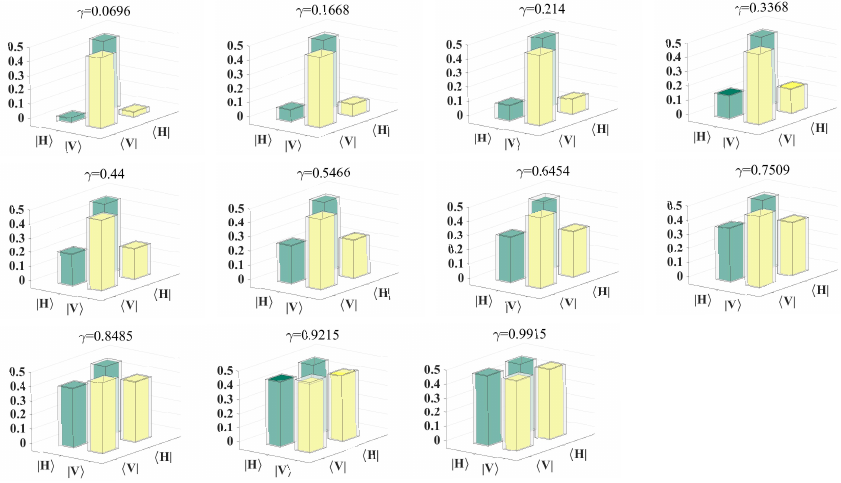}
        \caption{{\bf Quantum state tomography for the experimentally generated states $\rho_R^{\text{exp}}$.}
        The solid bars show the real parts of the tomographic states $\rho_R^{exp}(\gamma)$, whereas the transparent bars correspond to the theoretical values.
        }\label{Fig:tomoR}
\end{figure}

Then, $22.5^{\circ}$ oriented half-wave plate H3 prepares the state of the register $R$ into the state $1/\sqrt{2}(\ket{H}+\ket{V})$. After that the polarization of the photon is coupled by QP to its frequency distribution realizing a dephasing noise to vary $\gamma$ in $\rho_R(\gamma)$. Subsequently, the first VBD in FIG. \ref{fig:setup2}(b) directs the \textit{H} photon to the upper layer and \textit{V} photon to the lower layer to prepare the control state in the basis $\ket{u}$ and $\ket{l}$.
To analyze the specific form of the experimentally generated state $\rho_R^{exp}$, the optical axis of the H1 is horizontally placed to make all the photons pass through path "0". Then a standard quantum state tomography process is performed with a QWP and an HWP inserted before VBD, which behaves as a PBS now, to implement the three Pauli measurements. After the VBD, the photons are reflected out of the setup by a temporarily placed mirror and detected by the single-photon detectors. The detailed form of the tomographic state $\rho_R^{exp}$ can be found in FIG.~\ref{Fig:tomoR}.

For each experimentally generated state $\rho_R^{\text{exp}}$, we calculate its fidelity with the state $\rho_R(\gamma)$ given in Eq.~\eqref{eq:rhoR} for every $\gamma \in[0,1]$ (with the step length $10^{-4}$). Here the fidelity between quantum states $\rho$ and $\sigma$ is given by $F=\text{Tr} \left(\sqrt{\sqrt{\rho}\sigma\sqrt{\rho}}\right)$. We choose the $\gamma$ of the state $\rho_R(\gamma)$ which gives the highest fidelity and assign this value of $\gamma$ to the experimental state $\rho_R^{\text{exp}}$. In our experiment, the values of $\gamma$ and the corresponding thicknesses of the QP are given in TABLE~\ref{tab:d=2Table}. For every state $\rho_R^{\text{exp}}$, the corresponding fidelity is higher than $0.9995$.

After the preparation of the states in registers $B$ and $R$ a $22.5^{\circ}$ rotated half-wave plate H6 and two HBDs before and after it are used to implement the Hadamard operation, just as the HBD-HWP-HBD structure shown in FIG.~\ref{Fig:diagram}. Other wave plates H4, H5, and H7 are $45^{\circ}$ rotated altering the polarization of the photon to make the corresponding beams combine coherently in the right places.

At last, the second VBD converts the path DoF of the two layers to the polarization DoF and a QWP (Q1), a HWP (H8) and a PBS are used to perform a measurement on the register $R$ that aims to distinguish the quantum states $\rho_R^x$ in order for Bob to guess Alice's measurement outcome $X$. This is a projective measurement with POVM elements $\{\dyad{M_0}, \dyad{M_1}\}$ such that
\begin{equation}
\begin{aligned}
\ket{M_0}=\left(\sin\alpha\cos\beta+i\cos\alpha\sin\beta\right)\ket{0}
+\left(\cos\alpha\cos\beta-i\sin\alpha\sin\beta\right)\ket{0},\\
\ket{M_1}=\left(i\cos\alpha\cos\beta-\sin\alpha\sin\beta\right)\ket{0}
-\left(\cos\alpha\sin\beta+i\sin\alpha\cos\beta\right)\ket{1},
\end{aligned}
\label{Eq:ketM}
\end{equation}
where $\alpha=\theta_q$ and $\beta=\theta_q-2\theta_h$, and $\theta_q$ and $\theta_h$ are the rotated angles for QWP (Q1) and HWP (H8), respectively. We note that the measurement on the system $B$ takes place simultaneously with the measurement on $R$ and corresponds to the measurement of the path degree of freedom of the photon as shown in FIG. ~\ref{fig:setup2}. The settings of Q1 and H8 together with the corresponding guessing probabilities are also shown in TABLE~\ref{tab:d=2Table}.

\begin{table}
\centering
\begin{tabular}{cccccccccccc}

\hline
\hline
$\gamma$ & 0.9918 & 0.9221 & 0.8493 & 0.7509 & 0.6458 & 0.5466 & 0.4396 & 0.3369 & 0.2138 & 0.1662 & 0.0686 \\

n & 0 & $58\lambda$ & $97\lambda$ & $125\lambda$ & $154\lambda$ & $183\lambda$ & $209\lambda$ & $237\lambda$ & $266\lambda$ & $290\lambda$ & $305\lambda$\\
\hline
Q1($^{\circ}$)& -22.4 & -21.3 & -20.2 & -18.5 & -16.4 & -14.3 & -11.9 & -9.3 & -6.0 & -4.7 & -2.0 \\

H8($^{\circ}$)& 33.8 & 34.3 & 34.9 & 35.8 & 36.8 & 37.8 & 39.1 & 40.3 & 42.0 & 42.6 & 44.0\\
\hline
$P_{\text{guess}}^{\text{max}}$ & 0.9980 & 0.9809 & 0.9639 & 0.9421 & 0.9209 & 0.9029 & 0.8862 & 0.8731 &
0.8615 & 0.8584 & 0.8544 \\
$P_{\text{guess}}^{\text{exp}}$ & 0.9953 & 0.9776 & 0.9550 & 0.9301 & 0.9079 & 0.8891 & 0.8844 & 0.8702 & 0.8618 & 0.8610 & 0.8531 \\
& $\pm0.0003$ & $\pm0.0007$ & $\pm0.001$ & $\pm0.0012$ & $\pm0.0015$ & $\pm0.0016$ & $\pm0.0016$ & $\pm0.0017$ & $\pm0.0016$ & $\pm0.0017$ & $\pm0.0017$ \\
\hline
\hline
\end{tabular}
\caption{Settings and results for $d=2$ guessing game. The parameter $n$ gives the corresponding thickness  of the quartz plate used to generate state $\rho_R(\gamma)$, with $\lambda=808~\text{nm}$. The settings of the wave plates Q1 and H8, which are used to implement Bob's measurement (see FIG.~\ref{fig:setup2}(a)), are given in the middle lines. The last two lines $P_{\text{guess}}^{\text{max}}$ and $P_{\text{guess}}^{\text{exp}}$ give the optimal guessing probability predicted theoretically and the values obtained experimentally.}
\label{tab:d=2Table}
\end{table}

\subsection{Settings for the $d=3$ guessing game}
\label{sec:settingsd3}

For the $d=3$ guessing game, the input state $\rho_B$ is prepared by rotating the wave plates H1 and H2, and the phases between different path modes are generated by slightly tuning the first two HBDs. For our chosen relative phases the input state can be written as:
 \begin {equation}
 \ket{\psi}_B=e^{1.41i}\cos2\theta_1\cos2\theta_2\ket{0}-e^{1.66i}\cos2\theta_1\sin2\theta_2\ket{1}+\sin2\theta_1\ket{2}, \label{Eq:rhoB}
 \end {equation}
where $\theta_1$ and $\theta_2$ are the rotated angles for wave plates H1 and H2.

In the best known strategy, $\theta_1$ and $\theta_2$ are set to be $26.6^{\circ}$ and $5.9^{\circ}$, respectively. Moreover, we also test other input
states around the optimal one, and the detailed settings of $\theta_1$ and $\theta_2$ are shown in FIG.~\ref{fig:result2} and TABLE~\ref{tab:d=3Table}. In the $d=3$ guessing game, the corresponding optimal measurements used to distinguish states $\rho_R^x$ are performed using wave plates Q1 and H12, whose angles, together with the corresponding guessing probabilities, are also given in TABLE~\ref{tab:d=3Table}. The relation between the measurement basis and the angles of wave plates can be found in Eq.~\eqref{Eq:ketM}. Wave plates H5, H6, H8, and H11 are $45^{\circ}$ rotated to regulate the directions of the beams to make the photons combine coherently in the right places. The role of the remaining wave plates is discussed in Appendix~\ref{sec:FourierGate} in relation to the implementation of the three-dimensional Fourier gate.

\begin{table}
\centering
\begin{tabular}{ccccccccc}

\hline
\hline
state  & 1 & 2 & 3 & 4 & 5 & 6 & 7 & 8 \\
\hline
H1($^\circ$) & 22.6 & 24.6 & 26.6 & 28.6 & 30.6 & 26.6 & 26.6 & 26.6  \\
H2($^\circ$) & 5.9 & 5.9 & 5.9 & 5.9 & 5.9 & 1.9 & 9.9 & 17.9 \\
\hline
Q1($^\circ$) & 55.0 & 50.4 & 45.0 & -50.6 & -56.0 & -47.0 & 46.2 & 46.1 \\
H12($^\circ$) & 12.0 & 9.0 & 6.0 & 36.3 & 33.4 & 38.0 & 6.5 & 6.5 \\
\hline
$P_{\text{guess}}$ & 0.9669 & 0.9731 & 0.9753 & 0.9731 & 0.9664 & 0.9701 & 0.9702 & 0.9326 \\
$P_{\text{guess}}^{\text{exp}}$ & 0.9521 & 0.9466 & 0.9611 & 0.9628 & 0.9455 & 0.9419 & 0.9282 & 0.9281\\
 & $\pm0.0011$ & $\pm0.0011$ & $\pm0.001$ & $\pm0.0009$ & $\pm0.0011$ & $\pm0.0012$ & $\pm0.0012$ & $\pm0.0013$\\
\hline
\hline
\end{tabular}
\caption{
Settings and results for $d=3$ guessing game. Wave plates H1 and H2 are used to prepare the input states, and Q1 and H12 perform the corresponding optimal measurements, see the red and purple regions in FIG.~\ref{fig:setup2}(b). For each strategy, the theoretically predicted results are given as $P_{\text{guess}}$ and the corresponding experimental results as $P_{\text{guess}}^{\text{exp}}$.}
\label{tab:d=3Table}
\end{table}

\section{Implementation of the Fourier transformation operation}
\label{sec:FourierGate}

In the $d=3$ guessing game, the standard basis states $\ket{k}$ and the Fourier basis states
\begin{equation}
\ket{w_k}=U^{\dag}\ket{k}\, ,
\label{eq:FourierState}
\end{equation}
where $k=0, 1, 2$, constitute mutually unbiased bases and the Fourier transformation matrix is given by
\begin{equation}
 U=\frac{1}{\sqrt{3}}\left[
 \begin{matrix}
 1 & 1 & 1\\
 1 & w & w^2\\
 1 & w^2 & w
  \end{matrix}
  \right],
\end{equation}
with $w=e^{2\pi i/3}$.

Here we will show how to experimentally realize the transformation $U$. The method we use comes from Ref.~\cite{clements2016optimal}, which gives a universal algorithm to decompose such a multi-mode transformation matrix into a set of transformations $T_{m,n}$ between two modes $m$, $n$. Specifically, $U$ will be re-written as a sequentially ordered $T_{m,n}$, $U=D \prod_{(m,n)\in S}T_{m,n} $, where $S$ defines the order and D applied at the end adds an appropriate phase shift in each output mode. In the experiment, $T_{m,n}$ denotes a lossless variable beam splitter taking input modes $m$ and $n$, with the reflectivity $\cos\theta$ and phase shift $\phi$ at input $m$, where $\theta\in[0,\pi/2]$, $\phi\in[0, 2\pi]$:

\begin{equation}
T_{m,n}=\left[
 \begin{matrix}
   e^{i\phi}\cos\theta & -\sin\theta \\
   e^{i\phi}\sin\theta & \cos\theta
  \end{matrix}
  \right].
\end{equation}
Here we have omitted the nonfunctional elements of $T_{m,n}$.
This decomposition method is based on the work of Reck {\it et al.} \cite{reck1994}, and robust to the optical losses.

For our three mode transformation matrix $U$, we obtain the following expression according to the decomposition procedure in \cite{clements2016optimal}:

\begin{equation}
D=T_{1,2}T_{0,1}UT_{0,1}^{-1},
\end{equation}
which can be rewritten as $U=T_{0,1}^{-1}T_{1,2}^{-1}DT_{0,1}$. For any matrix $T_{m,n}^{-1}$, one can find a matrix $A_{m,n}$ and a matrix $D'$ so that $T_{m,n}^{-1}D=D'A_{m,n}$, then
\begin{equation}
U=D'A_{0,1}A_{1,2}T_{0,1},
\end{equation}
where for our $U$
\begin{equation}
\begin{aligned}
A_{0,1}&=\left[
 \begin{matrix}
 e^{\frac{-5i\pi}{6}}\cos(\frac{\pi}{4}) & -\sin(\frac{\pi}{4}) & 0 \\
 e^{\frac{-5i\pi}{6}}\sin(\frac{\pi}{4}) & \cos(\frac{\pi}{4}) & 0\\
 0 & 0 & 1
  \end{matrix}
  \right],  \\
A_{1,2}&=\left[
 \begin{matrix}
 1 & 0 & 0 \\
 0 & e^{\frac{2i\pi}{3}}\cos(\frac{54.74\pi}{180}) & -\sin(\frac{54.74\pi}{180}) \\
 0 & e^{\frac{2i\pi}{3}}\sin(\frac{54.74\pi}{180}) & \cos(\frac{54.74\pi}{180})
  \end{matrix}
  \right],  \\
T_{0,1}&=\left[
 \begin{matrix}
 e^{\frac{2i\pi}{3}}\cos(\frac{\pi}{4}) & -\sin(\frac{\pi}{4}) & 0 \\
 e^{\frac{2i\pi}{3}}\sin(\frac{\pi}{4}) & \cos(\frac{\pi}{4}) & 0 \\
 0 & 0 & 1
  \end{matrix}
  \right],  \\
D'&= \left[
 \begin{matrix}
 1&0&0\\
 0&e^{\frac{i\pi}{3}}&0\\
 0&0&e^{\frac{2i\pi}{3}}
 \end{matrix}
  \right].
\end{aligned}
\label{eq:AsandTs}
\end{equation}

\begin{figure}
\begin{center}
    	\includegraphics{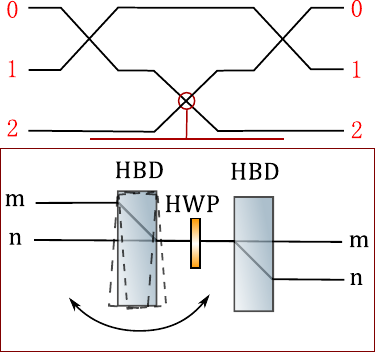}
        \caption{{\bf The structure diagram of the implementation of the Fourier transformation matrix.} In the top part, the lines represent the optical mode, and the crossing between two lines represents a variable beam splitter with a phase shift in one mode. In our experiment, we use two HBDs and an HWP, as shown in the bottom part of the figure, to realize the variable beam splitter, i.e., the crossing in the red circle, where the phase shift is added by slightly tuning the HBD. }\label{Fig:diagram}
    \end{center}
\end{figure}

In the following, we will show how to realize the variable beam splitter in our experiment. As the bottom part of FIG.~\ref{Fig:diagram} shows, an HBD-HWP-HBD structure is adopted to implement the $T_{m,n}$ ($A_{m,n}$). Firstly, an HBD maps the spacial path modes $\ket{m}$ and $\ket{n}$ in the input port into the polarization basis as follows: $\ket{m} \longrightarrow \ket{H}$ and $\ket{n} \longrightarrow \ket{V}$. Then the second HBD maps the polarization basis into the spacial path modes $\ket{m}$ and $\ket{n}$ again: $\ket{V} \longrightarrow \ket{m}$ and $\ket{H} \longrightarrow \ket{n}$. Hence, in the basis $\{\ket{H}, \ket{V} \}$, matrix $T_{m,n}$ ($A_{m,n}$) is represented as:

\begin{equation}
T_{m,n}^{\text{pol}}\left(A_{m,n}^{\text{pol}}\right)=\left[
 \begin{matrix}
 e^{i\phi}\sin\theta & \cos\theta \\
 e^{i\phi}\cos\theta & -\sin\theta
 \end{matrix}
   \right],  \\
\label{eq:TandAPol}
\end{equation}
where the superscript "pol" has been used to indicate the operation performed after conversion into the polarization encoding.
We now rewrite Eq.~\eqref{eq:TandAPol} such that
\begin{equation}
T_{m,n}^{\text{pol}} \left(A_{m,n}^{\text{pol}}\right)=   \left[
 \begin{matrix}
 \sin\theta & \cos\theta \\
 \cos\theta & -\sin\theta
 \end{matrix}
  \right]
  \left[
  \begin{matrix}
 e^{i\phi} & 0 \\
 0 & 1
  \end{matrix}
  \right].
 \end{equation}
Note that the first matrix:
\begin{equation}
R^{\text{pol}}=\left[
 \begin{matrix}
 \sin\theta & \cos\theta \\
 \cos\theta & -\sin\theta
 \end{matrix}
 \right]
\end{equation}
in the polarization basis can be implemented by a $(\pi/4-\theta/2)$ rotated HWP. The full $T_{m,n}$ ($A_{m,n}$) is realized by additionally applying phase shift by angle $\phi$ to the $\ket{H}$ basis state, which is realized by slightly tuning the first HBD.
In our experimental setup, the angles of middle-placed wave plates H7, H9 and H10 are $22.5^{\circ}$, $17.6^{\circ}$ and $22.5^{\circ}$ respectively.

The action of the matrix $D'$ does not affect Alice's measurement outcomes as her measurement is in the eigenbasis of $D'$.
However, it will contribute phase shifts to the post-measurement states of the register $R$: $\rho_R^x$.
For $x=0,1,2$, the phase shifts between the upper layer and lower layer are $0$, $e^{\frac{i\pi}{3}}$ and $e^{\frac{2i\pi}{3}}$, respectively. In our experimental setup, these phase shifts are added after the second VBD, where the upper layer and lower layer are translated to the polarization modes $V$ and $H$ respectively. Here, an individual wave plate with the rotated angle $0^{\circ}$ is inserted in the path $x$ to add the corresponding phase shift, which is not shown in the setup in FIG.~\ref{fig:setup2}.

We prepare the Fourier basis states $\ket{w_j}$ with $(j=0, 1, 2)$ defined in Eq.~\eqref{eq:FourierState} to probe the quality of the operation $U$. For the perfect gate the photon entering in state $\ket{w_j}$
 should be detected in the output mode $j$. The probabilities $P_{ij}^{\text{F,exp}}$ experimentally observed by obtaining a click in mode $i$ for input state $w_j$ are given in TABLE~\ref{tab:FourierTable}, from which we can see that the average probability of detecting the Fourier basis state in the correct mode is $0.9771\pm0.0009$, which shows the high quality of our Fourier gate implementation.

\begin{table}[htbp]
\centering
\begin{tabular}{c|ccc}

\hline
\hline
Probe state  & $\ket{w_0}$ & $\ket{w_1}$ & $\ket{w_2}$\\
\hline
Output mode 0 & $0.9722\pm0.001$ & $0.0176\pm0.0009$ & $0.0103\pm0.0007$ \\
\hline
Output mode 1 & $0.0183\pm0.0009$ & $0.9742\pm0.001$ & $0.0075\pm0.0006$ \\
\hline
Output mode 2 & $0.0095\pm0.0006$ & $0.0055\pm0.0005$ & $0.9851\pm0.0008$ \\
\hline
\hline
\end{tabular}
\caption{The normalized probabilities $P_{ij}^{\text{F,exp}}$ of the photon detected in mode $i$ for the probe state $\ket{w_j}$, where $i,j=0, 1, 2$. For each state, we collect about 20,000 photons in total. }\label{tab:FourierTable}
\end{table}

\section{Error analysis for the $d=2$ guessing game}
\label{sec:errord2}

For the $d=2$ guessing game, Bob should be able to perfectly guess Alice's measurement result $X$ if he has access to all the quantum information about Alice's measurement basis choice, i.e., the optimal guessing probability is $P^{\text{max}}_{\text{guess}}(\gamma=1,d=2)=1$. However, in our experiment we observe the highest value of $P^{\text{exp}}_{\text{guess}}(\gamma,d=2)=0.9953$. In the following, we provide a short numerical justification why the experimentally observed $P^{\text{exp}}_{\text{guess}}(\gamma,d=2)$ cannot reach $1$.

For that purpose we need to recall the two main sources of error in our experiment. The first one relates to the imperfections in the preparation of the state $\rho_R$. Specifically, using a QWP, an HWP and a PBS we perform a polarization analysis of the control state $\rho_R$ and estimate the highest experimentally achievable value of the coherence parameter to be $\gamma=0.9918$.

The second one relates to the imperfections of the interferometer.
The VBD transforms the photon's polarization degree of freedom to spacial modes $\ket{u}$ and $\ket{l}$. Ideally, after the photon undergoes the controlled Hadamard transformation $C_U=\dyad{0}\otimes\mathbb{I}_B+\dyad{1}\otimes H_B$, post-selecting on the measurement outcomes on the system $B$, we should obtain the following sub-normalized post-measurement states of the register R:
\begin{equation}
 \tilde{\rho}_R^x=\left[
 \begin{matrix}
 [\rho_R]_{00}\bra{x}\rho_B\ket{x}   & [\rho_R]_{01}\bra{x}\rho_BH^\dag\ket{x} \\
   [\rho_R]_{10}\bra{x}H\rho_B\ket{x} &   [\rho_R]_{11}\bra{x}H\rho_BH^\dag\ket{x}
  \end{matrix}
  \right],
\end{equation}
\label{Eq:post-measurest}
where $[\rho_R]_{ij}$ represents the matrix elements of the initial state of the register $R$ in the basis $\dyad{i}{j}$.
In our experiment of the $d=2$ game, the visibility of the interferometer composed of the two VBDs stays about $v=0.99$, which introduces a dephasing noise for the post-measurement state, such that:
\begin{equation}
\rho_{\text{out}}=\frac{1+v}{2}\rho_{\text{in}}+\frac{1-v}{2}\sigma_z\rho_{\text{in}}\sigma_z.
\label{Eq:DephNoise}
\end{equation}
Therefore, the state $\tilde{\rho}_R^x$ becomes
\begin{equation}
 \tilde{\rho}_R^{x, \text{deph}}=\left[
 \begin{matrix}
 [\rho_R]_{00}\bra{x}\rho_B\ket{x}   & v[\rho_R]_{01}\bra{x}\rho_BH^\dag\ket{x} \\
   v[\rho_R]_{10}\bra{x}H\rho_B\ket{x} &   [\rho_R]_{11}\bra{x}H\rho_BH^\dag\ket{x}
  \end{matrix}
  \right].
\end{equation}

Finally, Bob performs a measurement to distinguish the two possible states $\tilde{\rho}_R^{x, \text{deph}}$. In our experiment the measurements performed and the prepared input state $\rho_B$ are optimized for the ideal case, that is when the states to be distinguished are $\tilde{\rho}_R^{x}$ for the ideal initial state of $R$, namely
\begin{equation}
 \rho_R(\gamma)=1/2\left[
 \begin{matrix}
 1   & \gamma \\
   \gamma &  1
  \end{matrix}
  \right].
\end{equation}

For such ideal $d=2$ game the optimal input state $\rho_B$ for all $\gamma$ is the state $\ket{\psi}_B \propto \ket{0} + \ket{-}$ and for $\gamma = 0.9918$, the optimal measurement $\{M_0, M_1\}$ is given by:

\begin{equation}
M_0=\left[
 \begin{matrix}
 0.8550&0.3521 \\
    0.3521&0.1450
  \end{matrix}
  \right],
\,\,\,\,
 M_1=\left[
 \begin{matrix}
 0.1450 &  -0.3521\\
   -0.3521  &  0.8550
  \end{matrix}
  \right]
.
\end{equation}

This measurement is then applied to the actual state $\tilde{\rho}_R^{x, \text{deph}}$, where the actual initial state of the register $R$ prepared in the experiment is:

\begin{equation}
 \rho_R^{\text{exp}}=\left[
 \begin{matrix}
 0.5124   & 0.4955 - 0.0157i \\
   0.4955 + 0.0157i &  0.4876
  \end{matrix}
  \right].
  \label{eq:rhoRexp}
\end{equation}
The predicted detection probability $P_{ij}$ in output $D_{ij}$ is then given by
\begin{equation}
P_{ij}=\text{Tr}\left(M_i\tilde{\rho}_R^{x=j, \text{deph}}\right),
\end{equation}
where $i, j=0,1$. In this way, we calculate the probabilities in outputs $D_{00}$, $D_{01}$, $D_{10}$ and $D_{11}$ to be $0.5064$, $0.0024$, $0.0023$ and $0.4889$ respectively, and the estimated guessing probability $P_{\text{guess}}^{\text{est}}=P_{00}+P_{11}=0.9953$ agrees with the experimentally obtained probability $P_{\text{guess}}^{\text{exp}}=0.9953\pm0.0003$.

Moreover, by comparing the individual predicted outcomes with the values we obtained in the experiment shown in FIG.~\ref{fig:result1}(a), we can see that the probabilities in outputs $D_{00}$ and $D_{11}$ are consistent with the experimental values, and there is only a slight bias between the probabilities in outputs $D_{01}$ and $D_{10}$. Therefore, this noise model works well and the
errors for our $d=2$ guessing game are mainly coming from two imperfections, namely the preparation of the register state $R$, and the imperfect interference between the two layers.

\section{Error analysis for the $d=3$ guessing game}
\label{sec:errord3}
For the $d=3$ guessing game, we experimentally test the best known strategy and achieve a guessing probability $P_{\text{guess}}^{\text{exp}}(\gamma=0.9918, d=3)=0.9611\pm0.001$, which is indicated as data "3" in FIG.~\ref{fig:result2} and TABLE~\ref{tab:d=3Table}. In the following, we will justify this value by performing a detailed analysis of the experimental errors.

Besides the two main sources of error in the $d=2$ guessing game, i.e., the state preparation error for $\rho_R$ and the dephasing error between the layers $\ket{u}$ and $\ket{l}$, for the $d=3$ case the error occurring in the Fourier transform also needs to be included. Therefore, let us firstly discuss the main factors limiting the performance of our Fourier gate.

Let us now return to the setup used to test the quality of the Fourier gate discussed in Appendix~\ref{sec:FourierGate} and now depicted in FIG.~\ref{fig:Fouriernoise}. As described in Appendix~\ref{sec:FourierGate}, we implement the Fourier gate using a series of lossless variable beam splitters, which are denoted by the crossings between two modes in FIG.~\ref{fig:Fouriernoise}. The crossing A and C, B and D, and C and E constitute three M-Z (Mach-Zehnder) type interferometers, respectively. In our experiment of the $d=3$ game, the typical visibilities of all the interferometers are higher than $v = 0.98$.

We consider here the noise model described in~\cite{demkowicz2015}, where the imperfect visibility in the interferometer can be modeled through an additional fictitious mode $F$ which carries information about other degrees of freedom than the photon path. For perfect interference, $F$ is in the $\ket{0}_F$ state and is uncorrelated from the path information. However, imperfect interference can be seen as leakage of information into $F$, that is other degrees of freedom than photon path are then no longer the same for all the modes. It is the lack of access to $F$ which results in the effective decoherence of the qudit encoded in the photon path.

As an example let us examine first the imperfect interference between modes 0 and 1, which occurs e.g. on the crossing C in FIG.~\ref{fig:Fouriernoise}. According to the dephasing model in~\cite{demkowicz2015}, the mode mismatch between the interfering modes will lead to a correlated rotation of mode $F$. In other words, we can consider the input light modes traveling through fictitious beam splitters acting on mode $F$. These beam splitters split the input mode $\ket{0}_F$ into two orthogonal modes, where the transmitted part remains in state $\ket{0}_F$ while the reflected part is in the mode $\ket{1}_F$ for the signal being in mode $\ket{0}_B$ and it is in mode $\ket{2}_F$ for the signal being in mode $\ket{1}_B$. Therefore, the input state undergoes the following unitary transformation before the real beam splitter acting on system B:
\begin{equation}
\begin{aligned}
U_{BF}  &= \dyad{0}_B \otimes (\sqrt{v} \dyad{0}_F + \sqrt{1-v} \dyad{1}{0}_F - \sqrt{1-v} \dyad{0}{1}_F + \sqrt{v} \dyad{1}_F + \dyad{2}_F)   \\
        &+ \dyad{1}_B \otimes(\sqrt{v} \dyad{0}_F + \sqrt{1-v} \dyad{2}{0}_F - \sqrt{1-v} \dyad{0}{2}_F + \sqrt{v} \dyad{2}_F + \dyad{1}_F)+ \dyad{2}_B \otimes \id_F,
\end{aligned}
\end{equation}
where $v$ is the measured interferometric visibility in our $d=3$ guessing game. Since we do not have access to the register $F$, the state of the register $B$ before the real beam splitter can be described as:

\begin{equation}
    \rho_{\text{out}} = \text{Tr}_F[U_{BF} \rho_B \otimes \dyad{0}_F U_{BF}^\dag] = \sum_{i=0}^2 \bra{i}_F U_{BF} \ket{0}_F \rho_B (\bra{i}_F U_{BF} \ket{0}_F)^\dag.
\end{equation}

The resulting channel $\mathcal{K}_{0,1}$ describing the noise arising from the imperfect interference between modes 0 and 1 can be then written in the Kraus representation as follows:

\begin{equation}
    \begin{aligned}
    \mathcal{K}_{0,1}(\rho) &=\sum_{i=0}^2K_{i_{0,1}}\rho K_{i_{0,1}}^{\dag},\\
    K_{0_{0,1}} &= \bra{0}_F U_{BF}\ket{0}_F = \sqrt{v} (\dyad{0} + \dyad{1})_B + \dyad{2}_B, \\
    K_{1_{0,1}} &= \bra{1}_F U_{BF}\ket{0}_F = \sqrt{1-v} \dyad{0}_B, \\
    K_{2_{0,1}} &= \bra{2}_F U_{BF}\ket{0}_F = \sqrt{1-v} \dyad{1}_B. \\
    \end{aligned}
    \label{eq:channelKappa}
\end{equation}
Such a channel rescales the coherences between modes 0 and 1 by $v$ and all the coherences with mode 2 by $\sqrt{v}$.

\begin{figure}
\begin{center}
    	\includegraphics[width=0.8\columnwidth]{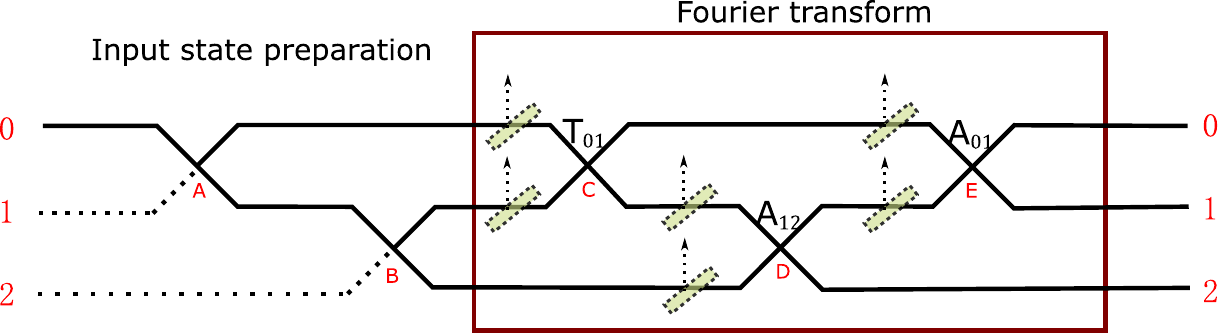}

        \caption{{\bf Illustration of the scenario used to estimate the quality of the Fourier transform.} A horizontal line represents a spacial mode, and a crossing between two lines represents a variable beam splitter with a phase shift in one mode. In our noise model, the dephasing errors can be modeled by adding the fictitious beam splitters before the crossings. }\label{fig:Fouriernoise}
    \end{center}
\end{figure}

Similarly, we can model the noise occurring when interfering modes 1 and 2 by the analogous channel $\mathcal{K}_{1,2}$. Hence, for the Fourier gate we implemented, the probability to detect a photon in output $i$ when inputting state $\ket{w_j}$ is given by

\begin{equation}
P_{ij}^{\text{F,deph}}=\bra{i}\mathcal{D'}\circ\mathcal{A}_{0,1}\circ \mathcal{K}_{0,1} \circ \mathcal{A}_{1,2}\circ\mathcal{K}_{1,2}\circ\mathcal{T}_{0,1}\circ \mathcal{K}_{0,1} (\ket{w_j}\bra{{w_j}})\ket{i},
\end{equation}
where $i,j=0, 1, 2$, and we define the channels corresponding to the operations $A$, $T$ and $D'$ defined in Eq.~\eqref{eq:AsandTs} as: $\mathcal{A}(\rho) = A\rho A^\dag$ and analogously for $\mathcal{T}$ and $\mathcal{D'}$. The obtained predicted values are shown in TABLE~\ref{tab:predTable}.

\begin{table}
\centering
\begin{tabular}{c|ccc}

\hline
\hline
Probe state  & $\ket{w_0}$ & $\ket{w_1}$ & $\ket{w_2}$\\
\hline
Output mode 0 & $0.9742$ & $0.0170$ & $0.0089$ \\
\hline
Output mode 1 & $0.0170$ & $0.9742$ & $0.0089$ \\
\hline
Output mode 2 & $0.0089$ & $0.0089$ & $0.9823$ \\
\hline
\hline
\end{tabular}
\caption{Predicted probabilities $P_{ij}^{\text{F,deph}}$ of obtaining a click in mode $i$ for each input state $\ket{w_j}$ according to our noise model. }\label{tab:predTable}
\end{table}

By comparing TABLE~\ref{tab:FourierTable} with TABLE~\ref{tab:predTable} we see that the corresponding probability distributions agree well which verifies that our analytical model provides a good description of the noise processes occurring in our experimental implementation of the three-dimensional Fourier gate.

In the implementation of our game only one of the two layers undergoes the Fourier operation. Let us then consider the corresponding noise model which includes the register $R$ and applies the noise to the state of register $B$ depending on the state of register $R$.  That is the channel $\mathcal{K}$ acts on part of the system $B$ correlated with the state $\ket{1}$ of the register $R$. The noise is then generated by the following unitary transformation acting on the extended space including register $F$:
\begin{equation}
    U_{RBF} = \dyad{0}_R \otimes \id_B \otimes \id_F + \dyad{1}_R \otimes U_{BF}.
\end{equation}

We can now calculate the Kraus operators of the channel $\mathcal{N}$ on the larger space $RB$ to get:
\begin{equation}
\begin{aligned}
    N_0 &=  \bra{0}_F U_{RBF}\ket{0}_F =\dyad{0}_R \otimes \id_B + \dyad{1}_R \otimes K_{0,B}, \\
    N_1 &=  \bra{1}_F U_{RBF}\ket{0}_F=\dyad{1}_R \otimes K_{1,B}, \\
    N_2 &=  \bra{2}_F U_{RBF}\ket{0}_F=\dyad{1}_R \otimes K_{2,B}.
\end{aligned}
\end{equation}
Note that we have omitted here the second level subscripts because the relation between Kraus operators $\{N_0, N_1, N_2\}$ and $\{K_0, K_1, K_2\}$ assumes the same form independently of which modes interfere.

Let us now consider a simple scenario in which we start with a product state $\rho_{R} \otimes \rho_{B}$ with $\rho_R$ given in Eq.~\eqref{eq:rhoR}. We then apply a single round of the channel $\mathcal{N}$  followed by the measurement of the system $B$. The sub-normalized state on $R$ conditioned on the outcome $x$ would then be:
\begin{equation}
\tilde{\rho}_R^x=\frac{1}{2}\left[
 \begin{matrix}
 \bra{x}\rho_B\ket{x}  & \gamma \bra{x}\rho_B K_0^{\dag}\ket{x} \\
 \gamma \bra{x}K_0\rho_B\ket{x} & \bra{x}\mathcal{K}(\rho_B)\ket{x} \\
  \end{matrix}
  \right].
 \end{equation}

Including all the noisy operations in the lower layer and the imperfections in the preparation of the initial state $\rho_R$, the actual final state conditioned on the outcome $x$ can be written as:

\begin{equation}
\tilde{\rho}_R^x=\left[
 \begin{matrix}
 [\rho_R]_{00}\bra{x}\rho_B\ket{x}  & [\rho_R]_{01}\bra{x}\rho_B (D'A_{0,1}K_{0_{0,1}} A_{1,2}K_{0_{1,2}}T_{0,1}K_{0_{0,1}})^{\dag}\ket{x} \\
   [\rho_R]_{10}\bra{x}D'A_{0,1}K_{0_{0,1}} A_{1,2}K_{0_{1,2}}T_{0,1}K_{0_{0,1}}\rho_B\ket{x} & [\rho_R]_{11}\bra{x}\mathcal{D'}\circ\mathcal{A}_{0,1}\circ \mathcal{K}_{0,1} \circ \mathcal{A}_{1,2}\circ\mathcal{K}_{1,2}\circ\mathcal{T}_{0,1}\circ \mathcal{K}_{0,1} (\rho_B)\ket{x} \\
  \end{matrix}
  \right],
 \end{equation}
 where $[\rho_R]_{ij}$ represents the matrix elements of the initial state of $R$, in our experiment given in Eq.~\eqref{eq:rhoRexp}, in the basis $\dyad{i}{j}$. Furthermore, recall from Eq.~\eqref{eq:channelKappa} that $\mathcal{K}_{0,1}(\rho)$ denotes a channel that rescales the coherences between modes 0 and 1 by $v$ and all the coherences with mode 2 by $\sqrt{v}$. The action of $\mathcal{K}_{1,2}(\rho)$ is analogous. Also recall that $K_{0_{0,1}}$ is the $K_0$ Kraus operator which is a diagonal matrix with $\sqrt{v}$ in the first two diagonal entries and 1 in the third one. The structure of $K_{0_{1,2}}$ is analogous.

Finally, we also need to include the dephasing noise between the two layers by rescaling the two off-diagonal entries by a factor $v$ after the channel:

\begin{equation}
 \tilde{\rho}_R^{x, \text{deph}}=\frac{1+v}{2}\tilde{\rho}_R^x+\frac{1-v}{2}\sigma_z\tilde{\rho}_R^x\sigma_z,
\end{equation}
similarly to Eq.~\eqref{Eq:DephNoise}.

For data "3" in FIG.~\ref{fig:result2}, we test the best known strategy for the ideal register state $\rho_R(\gamma=0.9918)$, and the optimal input state is given by $\ket{\psi}_B=a_1\ket{0}+a_2\ket{1}+a_3\ket{2}$ with the coefficients $a_1=0.0938 + 0.5786i$, $a_2=0.0109 - 0.1218i$ and $a_3=0.8009$. Now we can predict the detection probability $P_{ij}$ in output $D_{ij}$ as
\begin{equation}
 P_{ij}=\text{Tr}\left(M_i\tilde{\rho}_R^{x=j, \text{deph}}\right),
\end{equation}
where $i,j=0,1,2$ and $\{M_0, M_1, M_2\}$ with
\begin{equation}
M_0=\left[
 \begin{matrix}
 0.5003&0.2027 + 0.4571i \\
 0.2027 - 0.4571i&0.4997
  \end{matrix}
  \right],
  \,\,\,\,
  M_{1} = \mathbf{0} ,
\,\,\,\,
 M_2=\left[
 \begin{matrix}
 0.4997 &  -0.2027-0.4571i\\
 -0.2027+0.4571i  &0.5003
  \end{matrix}
  \right]
\end{equation}
is the projective POVM measurement performed by Bob to guess Alice's outcome $x$.

Here we need to mention that for the optimal strategy in the $d=3$ game, the projective measurements are sufficient, i.e., we only aim to distinguish the two dominant outcomes of the three outcomes on system $B$. In our scenario, the measurements $M_0$ and $M_2$ allow us to distinguish the states $\rho_R^0$ and $\rho_R^2$, respectively. Then we can estimate the guessing probability $P_{\text{guess}}=P_{00}+P_{22}=0.9554$ for $v = 0.98$. With the experimentally observed value $P_{\text{guess}}^{\text{exp}}(\gamma=0.9918, d=3)=0.9611\pm0.001$ and taking into account the fact that the actual visibilities can be slightly higher than $v=0.98$, we can see that the proposed model provides a good description of the noise processes occurring in the experiment. Specifically, due to a large number of interferometers for the $d=3$ game, we see that the imperfect visibility has a significant impact on the observed guessing probability.

Moreover, when considering the other data points shown in FIG.~\ref{fig:result2}, we also need to include an additional error source. In our experiment, the phases of the interferometers are calibrated to prepare the input state for the data "3" to implement the best known strategy, and then other strategies are probed by varying the angles of H1 and H2. Since the surface of the wave plate is not absolutely smooth, the phase of the interferometer will undergo small change while rotating the wave plate. This has a significant effect, especially for H2, for which the photons in the two arms of the interferometer pass through two different places. Then the surface irregularity of the wave plate introduces a relative phase in the prepared state, see FIG.~\ref{fig:setup2}. That is also the reason why data points "6" and "7" have larger deviations from the corresponding theoretical values. Therefore, besides the error sources we mentioned above, the error in the preparation of $\rho_B$ should also be included for data points other than data "3".

\end{document}